\documentclass[%
 reprint,
superscriptaddress,
 amsmath,amssymb,
 aps,
]{revtex4-2}

\usepackage{graphicx}
\usepackage{dcolumn}
\usepackage{bm}

\usepackage[dvipsnames]{xcolor}
\usepackage{stackrel}
\usepackage{sidecap} 

\usepackage{amsfonts}
\usepackage[normalem]{ulem}

\usepackage{tabularx}

\newcommand{\dd}{\mathrm{d}}
\newcommand{\dt}[1]{\frac{\dd #1}{\dd t}}
\newcommand{\dtn}[1]{\frac{\dd #1}{\dd \tau}}
\newcommand{\cb}{X}
\newcommand{\cc}{S}
\newcommand{\ccd}{C}
\newcommand{\cbn}{x}
\newcommand{\ccn}{s}

\newcommand{\cco}{K_S}
\newcommand{\ccdo}{K_C}
\newcommand{\cw}{W}
\newcommand{\tauc}{\tau_{\rm food}}
\newcommand{\taud}{\tau_{\rm death}}
\newcommand{\death}{\lambda}
\newcommand{\influx}{n}
\newcommand{\outflux}{n_{\rm out}}

\begin{document}

\preprint{APS/123-QED}


\title{How adaptation to food resources and death rates shape \\ oscillatory dynamics in a microbial population}


\author{Benedetta Ciarmoli}
\affiliation{CNRS, Sorbonne Universit\'{e}, Physicochimie des Electrolytes et Nanosyst\`{e}mes Interfaciaux, F-75005 Paris, France}

\author{Sophie Marbach}
\email{sophie.marbach@cnrs.fr}
\affiliation{CNRS, Sorbonne Universit\'{e}, Physicochimie des Electrolytes et Nanosyst\`{e}mes Interfaciaux, F-75005 Paris, France}

\date{\today}

\begin{abstract}
Microbes constantly interact with their environment by depleting and transforming food sources. Theoretical studies have mainly focused on Lotka-Volterra models, which do not account for food source dynamics. In contrast, consumer-resource models, which consider food source dynamics, are less explored. In particular, it is still unclear what physical mechanisms control oscillatory dynamics at a single population level, a phenomenon which can only be captured by a consumer-resource model. 
Here, we present a minimalistic consumer-resource model of a single microbial population with growth and death dynamics, consuming a continuously replenishing substrate. 
Our model reveals that decaying oscillations can occur around steady state if and only if the timescale of microbial adaptation to food supply changes exceeds the death timescale. This interplay of timescales allows us to rationalize the emergence of oscillatory dynamics when adding various biophysical ingredients to the model.
We find that microbial necromass recycling or complementary use of multiple food sources reduces the parameter range for oscillations and increases the decay rate of oscillations. Requiring multiple simultaneous food sources has the opposite effect. Essentially, facilitating growth reduces the likelihood of oscillations around a fixed point. We further demonstrate that such damped oscillatory behavior is correlated with persistent oscillatory behavior in a noisy environment.  
We hope our work will motivate further investigations of consumer-resource models to improve descriptions of environments where food source distributions vary in space and time. 
\end{abstract}

\maketitle



Microbes are at the root of diverse and essential physico-chemical reactions. In soils, as a canonical example, microbial decomposition is a first step towards carbon sequestration~\cite{walling2020review,de2021closed,amelung2020towards,lehmann2020persistence,schloter2018microbial}. Microbes also act as water filters, pollutant removers, promote plant growth, and participate in the fixation of numerous chemical compounds such as nitrogen~\cite{de2021closed,amelung2020towards,lehmann2020persistence,schloter2018microbial,nannipieri2017microbial,gowda2022genomic}. Beyond soils, microbes play key roles in carbon cycling in the ocean~\cite{datta2016microbial,nguyen2022microbes}. From a biomedical perspective, a canonical example is their role on healthy gut function~\cite{waclawikova2022gut}. From an industrial perspective, microbial metabolism is also useful for the large scale bioproduction of chemicals~\cite{giri2020harnessing}. 

In this context, one must understand the impact of food source abundance on microbial density and function, and reciprocally how microbes affect food source abundance. For instance, slight variations in local oxygen concentration can modify the emission of methane by microbes~\cite{angle2017methanogenesis}. In soils, 
there are many nutrients whose cycle is not investigated which would help us understand community diversity and soil function~\cite{nannipieri2017microbial}. From an industrial perspective, there are still open challenges in maximizing the efficiency of bioproduction or detoxification by tuning \textit{e.g.} nutrient or toxin supply conditions~\cite{giri2020harnessing,shibasaki2020controlling}.
The balance between the relative abundance of food sources and the living microbes is thus subtle, calling for fundamental insights. 


From a theoretical perspective, most efforts have focused on microbe-level population dynamics, described via Lotka Volterra or Macarthur models and  derivatives~\cite{macarthur1970species,hatton2024diversity,hu2022emergent,bunin2017ecological,altieri2021properties,roy2020complex,may2007theoretical,deng2022dynamic}. In these models, all species are biotic, \textit{i.e.} living and reproducing. While fundamental to predict and understand biodiversity in microbial communities, such theories do not account for the underlying abundance of food sources and interactions mediated by these relative abundances. In this context, food sources refer to abiotic, non-living compounds. This means that under different initial food source conditions, LV type models may not be able to predict correct microbial dynamics~\cite{picot2023microbial,o2018whence}.  Phenomena like resilience, niche expansion, and spatial self-organisation cannot be accounted for without considering local couplings and/or underlying food source dynamics~\cite{van2022ecological,dal2019emergent}. Finally, LV-type models can not inform on nutrient cycling dynamics. 

In contrast, consumer-resource (CR) models, where food sources are governed by abiotic growth laws, have allowed us to progress on microbe mediated nutrient cycling~\cite{chesson1990macarthur,picot2023microbial}. 
Unlike the Macarthur model~\cite{macarthur1970species}, which is puzzlingly also called ``consumer-resource model'', here the limiting food source for microbial growth is an abiotic (non-reproducing) species~\cite{chesson1990macarthur,picot2023microbial}. CR models have investigated the role of switching environment parameters or toxin levels on population diversity and growth~\cite{shibasaki2021exclusion,meacock2025environment}. Comparing CR models to experiments has allowed us to explain denitrification rates in large communities~\cite{gowda2022genomic}, community diversity~\cite{goldford2018emergent}, oscillations in planktonic populations~\cite{huisman1999biodiversity} and degradation rates in the ocean~\cite{nguyen2022microbes}. Even at larger scales, accounting explicitly for food sources is key to understand pattern arrangements of for instance sessile organisms like trees~\cite{lee2021growth}. Yet, in comparison with LV models, there is still much less fundamental insight on the variety of dynamics that CR models can lead to~\cite{o2018whence}.

\begin{figure*}
    \centering
    \includegraphics[width=0.7\textwidth]{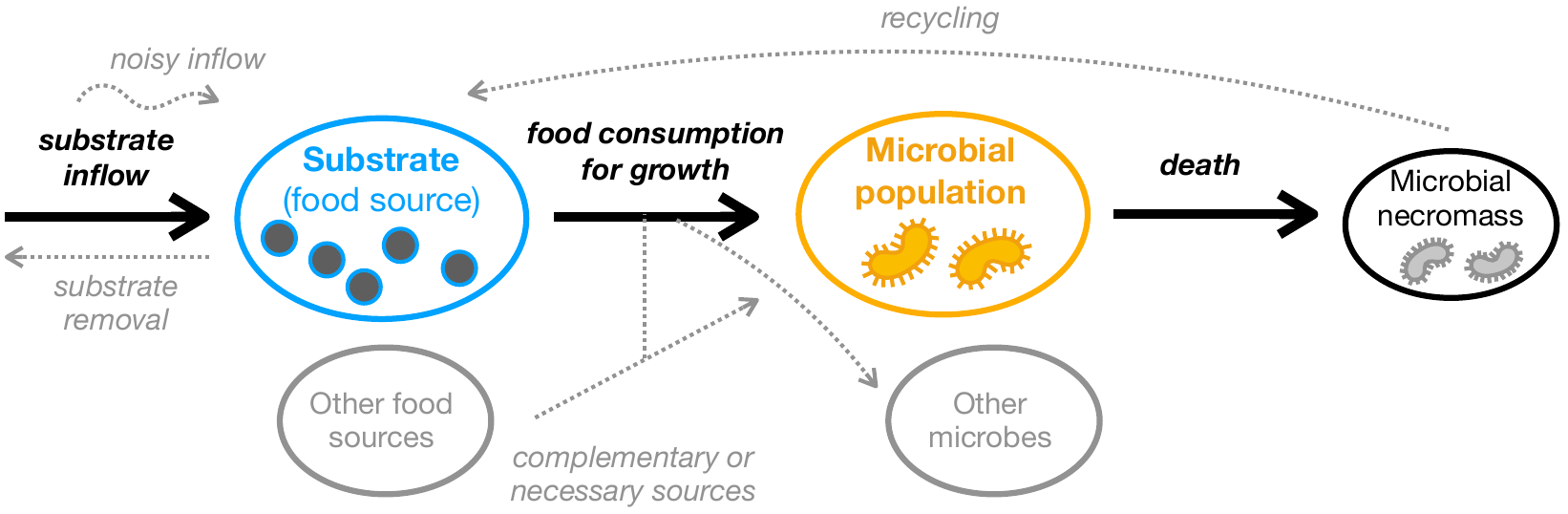}
    \caption{\textbf{Schematic of the minimal consumer-resource (CR) model considered in this study} involving a single abiotic resource -- the substrate in blue -- and a single consumer -- the microbial population in yellow. Black arrows indicate the main physical processes at play. Here, we study the emergence of damped oscillatory dynamics in this minimal model and several model extensions corresponding to further bio-physical ingredients in gray. }
    \label{fig:figure1}
\end{figure*}

One fascinating fundamental feature of CR models is the possibility of observing oscillatory dynamics on the abundance of one or more species caused by oscillations of food source content~\cite{fowler2014oscillations,fowler2014starvation,huisman1999biodiversity}.
Strikingly, even in the simplest CR setting, with 1 species and 1 limiting food source, under low food source supply rates, the abundance of microbes can undergo decaying oscillatory dynamics during the relaxation from a perturbation at steady state~\cite{khatri2012oscillating}. For a single species, LV dynamics does not predict any oscillatory behavior, even damped. In fact, steady-state oscillations of microbial population are quite universal for LV dynamics, but they involve always at least 2 reproducing species. Since only few works have higlighted the presence of decaying oscillations for a single species, theoretically~\cite{khatri2012oscillating,skichko2006mathematical}, or in continuous culture experiments~\cite{borzani1977some,porro1988oscillations}, it is still not clear what minimal ingredients are required for oscillations of a single microbe species, or how these oscillations are modified by diverse physical phenomena. We lack physical intuition on the underlying mechanism which determine oscillation features: the parameter range over which they occur and the decay rate which sets how fast the system reaches equilibrium.

Here we investigate a simplistic model of a consumer growing on a single food source which exhibits transient oscillatory dynamics -- see Fig.~\ref{fig:figure1} -- and perform analytical and numerical investigations of the model. We find decaying oscillations occur around steady states for a certain range of parameters. Oscillations also increase the time to reach a stable steady state compared to non-oscillatory solutions. We also demonstrated that in regions of the parameter where damped oscillations occur in the deterministic case, persistent oscillations emerge in the presence of a noisy inflow of food sources (Sec. I). From this starting point, we consider the impact of further biophysical features: recycling of microbial necromass (Sec. II), multiple resources being either necessary or complementary for growth (Sec. III), and further environmental factors such as the presence of multiple species or substrate removal (Sec. IV). Our main finding is that in spite of these biological differences, the behavior of the system can always be captured by a simple ratio of timescales. Indeed, the emergence of damped oscillatory behavior, and the magnitude of the decay rate of these oscillations, are characterized by a balance of only two timescales: the timescale to adapt to changes in food abundance $\tau_{\rm food}$, and the timescale to recover from excess population numbers, from death or predators, $\tau_{\rm death}$. When $\tau_{\rm food}  \leq \tau_{\rm death}$, the system adapts sufficiently rapidly to changes in food supply rates that no oscillations are seen and the system converges rapidly to steady state. Reciprocally, when $\tau_{\rm food}  \geq \tau_{\rm death}$, these mechanisms become out of phase, oscillations occur, and the system converges more slowly towards its steady state. Changes to the model modify these timescales and thus the balance of competing phenomena and the dynamics. In particular, phenomena that facilitate growth such as recycling or complementary food sources decrease the time to respond to changes in food supply, reduce the range over which oscillations occur and the timescale for oscillations to relax.

\section{General framework to study oscillations in a single microbial population}

\subsection{Consumer-resource model}

Our starting point is a consumer-resource model with minimal ingredients. We let $X(t)$ and $S(t)$ be the microbial population and food source -- sometimes called substrate -- concentration, respectively, with time $t$. Throughout this work, we do not specify precisely the units of all quantities, but we note that the units of $X(t)$ and $S(t)$ may be chosen with no change in the equations -- $X(t)$ can therefore be measured \textit{e.g.} in optical units as common for fluorescent species and $S(t)$ in $\mathrm{mol/L}$. The food source can represent, for instance, a carbon source -- see Fig.~\ref{fig:figure1}. 
Since the model should represent microorganisms, they should at least consume food sources that allow them to grow and reproduce. 
Then, we require two necessary conditions to reach non-trivial steady states. First, food sources should be constantly supplied through a substrate inflow, otherwise microorganisms would run out of supplies. This supply could originate from incoming flows in their environment. Second, there must be a mechanism by which the population of microorganisms diminishes, otherwise the population would indefinitely grow. This may be due to external conditions like flows which carry microorganisms away from the considered pool, external predators, or death due to external stresses \cite{camenzind2023formation}. The kinetics of a single species $X(t)$ and its food source $S(t)$ are thus minimally described by
\begin{align}
    \dt{\cb} &= r \mu(\cc) \cb - \death \, \cb \label{eq:model1bact} \\
    \dt{\cc} &= \influx \cco -  \gamma  r \mu(\cc)\cb \label{eq:model1substrate}
\end{align} 
where $\mu(\cc)$ is nondimensional and quantifies the efficiency of reproduction depending on the amount of available food sources, $r$ is a reproduction rate, $\death$ is a death rate, $\influx$ is an inflow rate relative to a reference concentration $\cco$ of supplied substrate, and $\gamma$ is a conversion factor quantifying the amount of substrate needed per microorganism to sustain reproduction and growth. 

\begin{figure}
    \centering
    \includegraphics[width=0.99\linewidth]{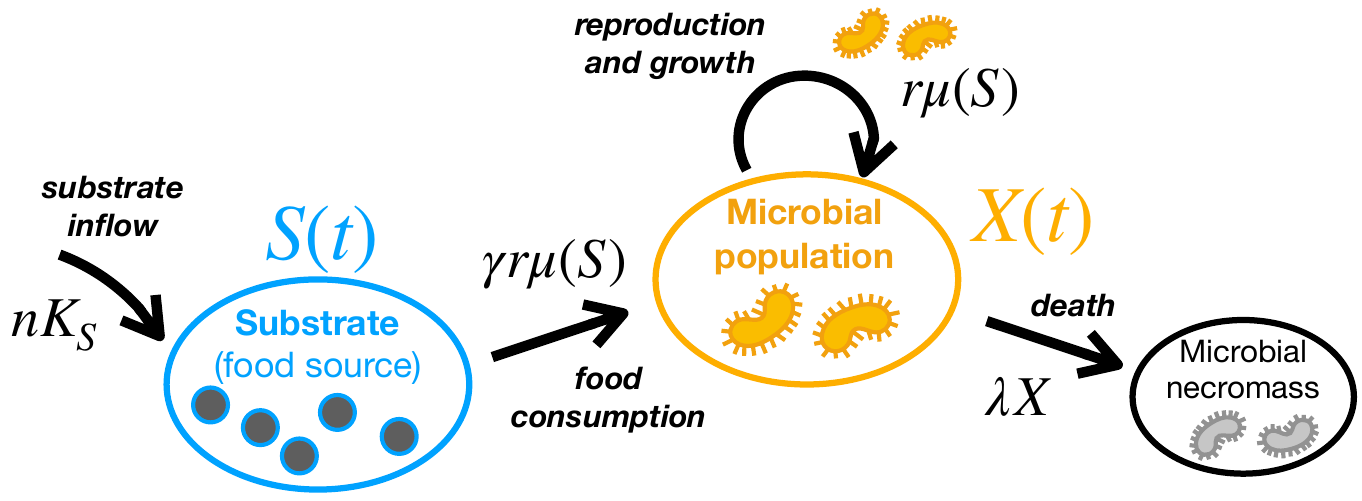}
    \caption{\textbf{Schematic of the minimal reference model in Eqs.~(1-2)}.}
    \label{fig:figure2}
\end{figure}

The ingredients of the model described by Eqs.~(1-2) and illustrated in Fig.~\ref{fig:figure2} are generic and broadly used, albeit often with additional terms~\cite{koffel2021competition,o2018whence,reynolds2013can,picot2023microbial,khatri2012oscillating}. In particular, these modeling ingredients have been shown to describe well the response of a heterogeneous system under various conditions, so they are a priori a good coarse-graining of the underlying microscopic dynamics~\cite{welker2024dividing}. 

A few aspects of the model merit discussion, for the sake of generality. 
In practice, one could assume Monod kinetics for substrate consumption, as is common, as $\mu(\cc) = \frac{\cc}{\cco + \cc}$, see \textit{e.g.} Refs.~\cite{monod2012growth,fowler2014oscillations,fowler2014starvation,khatri2012oscillating,de2021closed}. For now, however, we keep the derivation general and assume $\mu(\cc)$ can take any form so long as it is a continuous function of $\cc$. This function, in our minimal model, should nonetheless be increasing as one expects that more resources should favor growth. 
Different authors place $\gamma$ either in the species equation~\cite{de2021closed} or, like here, in the substrate's equation~\cite{khatri2012oscillating}, and thus $\gamma$ can assume very different values depending on the definition. Note that we do not assume equal rates of substrate and biomass removal from the system, which would be the case to model a chemostat growth setup~\cite{picot2023microbial,khatri2012oscillating} but we will review the impact of substrate removal in Sec.~\ref{sec:substrateremoval}. However, Eqs. (1-2) represent well systems where the inflow of some food sources is different from the dilution rate of the chemostat. This is the case for instance for oxygen supply, since it is often provided via a gazeous pump, and in this context, single species oscillations have been observed~\cite{borzani1977some,porro1988oscillations}. 

Our goal in this paper is to study the minimal model described by Eqs.~(1-2) and several extensions, so as to understand how and why oscillations might arise in the microbial and substrate concentrations, how they might be suppressed or enhanced, and how their temporal features, decay rate and oscillation frequency, depend on parameters.

\subsection{Non-dimensionalization and essential system parameters.}

To make progress, it is useful to notice a natural non-dimensionalization of variables as $$ \cbn = \frac{\cb}{\cco/\gamma}, \, \ccn = \frac{\cc}{\cco} \,\, \text{and} \,\, \tau = \death \, t $$ such that $\cco$ sets the scale of the substrate concentration, $\cco/\gamma$ that of the microbial population, and the death rate $\death$ is a natural timescale. The equations become
\begin{equation}
\begin{split}
    \dtn{\cbn} &= \frac{r}{\death} \mu(\ccn) \cbn -  \cbn  \\
    \dtn{\ccn} &= \frac{\influx}{\death} - \frac{r}{\death} \mu(\ccn) \cbn 
    \label{eq:framework}
    \end{split}
\end{equation}
The only remaining relevant parameters are thus the non-dimensional reproduction rate $r/\death$ and the substrate supply $\influx/\death$. We must thus learn about dynamics within this phase diagram.

\begin{figure*}
    \centering
    \includegraphics[width=0.99\textwidth]{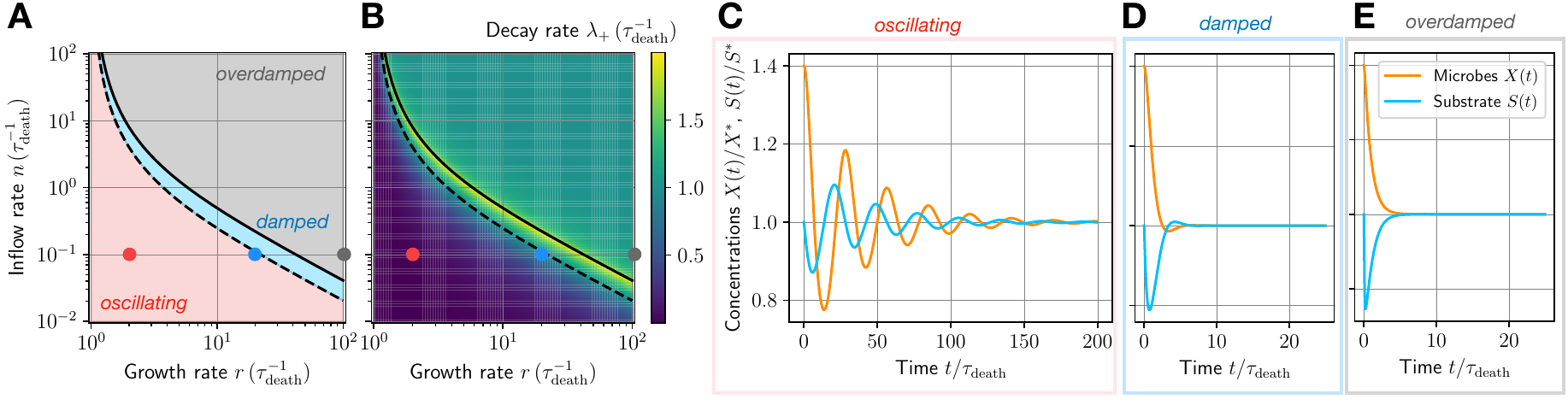}
    \caption{\textbf{Possible dynamical behaviors in the reference model.} (A) phase diagram representing the possible dynamics in the inflow rate versus growth rate space. (B) Same diagram but with colors indicating the magnitude of the damping rate. In both (A,B) the full line indicates the limit $\beta = 4$ and the dashed line the limit $\beta = 2$. (C) Typical oscillating solution, using $r = 2 \ \taud^{-1}$ and $n = 0.1 \ \taud^{-1}$; (D) damped solution using $r = 20 \ \taud^{-1}$ and $n = 0.1 \ \taud^{-1}$ and (E) overdamped solution, using $r = 100 \ \taud^{-1}$ and $n = 0.1 \ \taud^{-1}$. The locations of (C, D, E) are indicated in (A-B) with colored dots. The legend is shared across (C, D, E).}
    \label{fig:figure3}
\end{figure*}

\subsection{Fixed points, decay rates, and oscillating behavior}

\paragraph*{Fixed point $(\cb^*, \cc^*)$.} The only fixed point of the system is given by searching for $\dt{\cb} = \dt{\cc} = 0$ and so 
\begin{equation}
    (\cbn^*,\ccn^*) = \left( \frac{\influx}{\death}, \mu^{-1} \left( \frac{\death}{r}\right) \right) 
    \label{eq:fixedGeneral}
\end{equation}
where $\mu^{-1}(\ccn)$ is the inverse function of $\mu(\ccn)$, which is also an increasing function of $\ccn$. 
Quite naturally, the death rate $\death$ decreases the equilibrium microbial concentration, and increases the equilibrium food concentration. Also naturally, a higher food supply rate $n$ increases the equilibrium microbial population $\cb^*$, yet it is surprising that it does not set the equilibrium food concentration. Instead, the equilibrium food concentration is given by a balance of two rates: microbe consumption rate and death rate. 

One can note that there also exists a trivial stationary point where the population is extinct $\cb^* = 0$, and the substrate concentration increases at a constant rate according to the substrate inflow $\dt{\cc} = n \cco$. 

\vspace{2mm}

\paragraph*{Decay rates.} Around the fixed point, one may evaluate the stability of the system by investigating the Jacobian of the system. Formally if we define $\mathcal{F}_\ccn$ and $\mathcal{F}_\cbn$ as $\dt{\cbn} = \mathcal{F}_\ccn(\cbn,\ccn)$ and $\dt{\ccn} = \mathcal{F}_\cbn(\cbn,\ccn)$, then the response of the system near a fixed point $(\cbn,\ccn) = (\cbn^* + \delta \cbn, \ccn^* + \delta \ccn)$ is given by
\begin{equation}
    \begin{pmatrix}
        \dt{\delta \cbn} \\  \dt{\delta \ccn} 
    \end{pmatrix} = J \begin{pmatrix}
      {\delta \cbn} \\  {\delta \ccn} 
    \end{pmatrix}, \,\, \text{with} \,\, J = \begin{pmatrix}
        \partial_\cbn \mathcal{F}_\cbn & \partial_\ccn \mathcal{F}_\cbn \\
        \partial_\cbn \mathcal{F}_\ccn & \partial_\ccn \mathcal{F}_\ccn
    \end{pmatrix}
\end{equation}
and each function is evaluated at the fixed point. Here,
\begin{equation}
    J = \begin{pmatrix}
        0 &  \frac{r}{\death} \mu'(\ccn^*)\cbn^* \\
        -\frac{r}{\death} \mu(\ccn^*) & - \frac{r}{\death} \mu'(\ccn^*)\cbn^*
    \end{pmatrix} =  \begin{pmatrix}
        0 & \beta \\
         - 1 & - \beta 
    \end{pmatrix}
\end{equation}
where we have defined the non-dimensional parameter 
\begin{equation}
    \beta  =  \frac{r \mu'(\ccn^*)\cbn^*}{\death} 
    \label{eq:beta}
\end{equation}
whose interpretation we come back to in the following section. Let $\tilde{\lambda}_\pm = \lambda_{\pm}/\lambda$ be the nondimensional eigenvalues of this nondimensional matrix, which provide an information on decay rates. They are given by 
\begin{equation}
    \frac{\lambda_{\pm}}{\death} = \frac{-\beta \pm \sqrt{\beta^2  - 4 \beta}}{2}.
\end{equation}
The eigenvalues are either (i) real and negative or (ii) complex with a negative real part, corresponding to decaying solutions towards the fixed point either in an (i) overdamped or (ii) oscillating way. The oscillations -- in both microbial population and food concentration -- occur if $\beta^2 \leq 4 \beta$ or equivalently $\beta \leq 4$ which translates into a constraint on $\influx$ and $r$. 
These oscillations occur with frequency 
\begin{equation}
    \frac{\Omega}{\death} = \frac{\sqrt{4 \beta - \beta^2}}{2}. 
    \label{eq:frequency}
\end{equation}
Significant oscillations can only be seen when the oscillation frequency $\Omega$ is faster than the damping rate (given by the real part of $\lambda_{\pm}$) and so only when $\beta \leq 2$. In the intermediate regime where $2 \leq \beta \leq 4$, the damping rate is so strong that oscillations do not have time to establish, resulting in a ``damped'' regime. For $\beta \geq 4$, no oscillations can arise resulting in ``overdamped'' dynamics. 

So far, we have simply assumed $\mu(s)$ to be a continuous, increasing function of $s$. Given the steady state values, we can see that $\beta = r n \mu'(\mu^{-1}(\death/r))/\death^2$, and hence if the resource input flow $n$ is sufficiently small, there will exist a range of values for $r$ where $\beta \leq 4$ and oscillations should arise. It is not necessary to assume more properties of $\mu(s)$ for oscillations to emerge. Oscillations are thus a quite generic feature of a single species model. 

To illustrate these mathematical results, we take as an example the case of Monod kinetics where $\mu(s) = s/(1+s)$. We reproduce the condition for oscillations or not in Fig.~\ref{fig:figure3}-A in a phase diagram depending on the only 2 relevant physical parameters: the non-dimensional reproduction rate $r/\death$ and the substrate supply $\influx/\death$. We find that oscillations predominantly occur for sufficiently small inflow rates, as was also seen in a closely related model~\cite{khatri2012oscillating}. In addition, in the oscillating region, the decay rate of the oscillations, $\beta/2$, is, most of the time, much smaller than in the overdamped case, where it is given by the smallest eigenvalue (in amplitude), $\lambda_+$, see Fig.~\ref{fig:figure3}-B. In the intermediate ``damped'' regime, we find the decay rate is quite large, preventing dynamics from having the time to exhibit oscillations. We present typical examples of an oscillating solution in Fig.~\ref{fig:figure3}-C, a damped in Fig.~\ref{fig:figure3}-D and an overdamped one in Fig.~\ref{fig:figure3}-E. Clearly, the damped and overdamped solution converge much faster to the stationary state.

\subsection{Mechanism for the emergence of oscillating solutions}

Now we interpret these mathematical results in mechanistic terms. Firstly, why do the oscillations appear? To interpret this, we closely follow Ref.~\cite{khatri2012oscillating}. Suppose that as a starting point, the biomass is slightly in excess compared to the fixed point, but the substrate concentration is given by its steady state value $\cc^*$. This means that the rate of substrate consumption exceeds the inflow rate: thus the substrate concentration decreases. As a result, the substrate concentration quickly reaches a regime in which it cannot sustain microbial growth, and the biomass concentration decreases. At some point, biomass concentration has decreased sufficiently that again the substrate inflow rate exceeds the consumption rate: substrate concentration increases again but is still too low to allow for biomass growth. Eventually, the substrate concentration is high enough to allow for biomass growth again, until the biomass exceeds a certain threshold, where the substrate concentration decreases again, and the cycle starts again.

\begin{figure}
    \centering
    \includegraphics[width=0.99\linewidth]{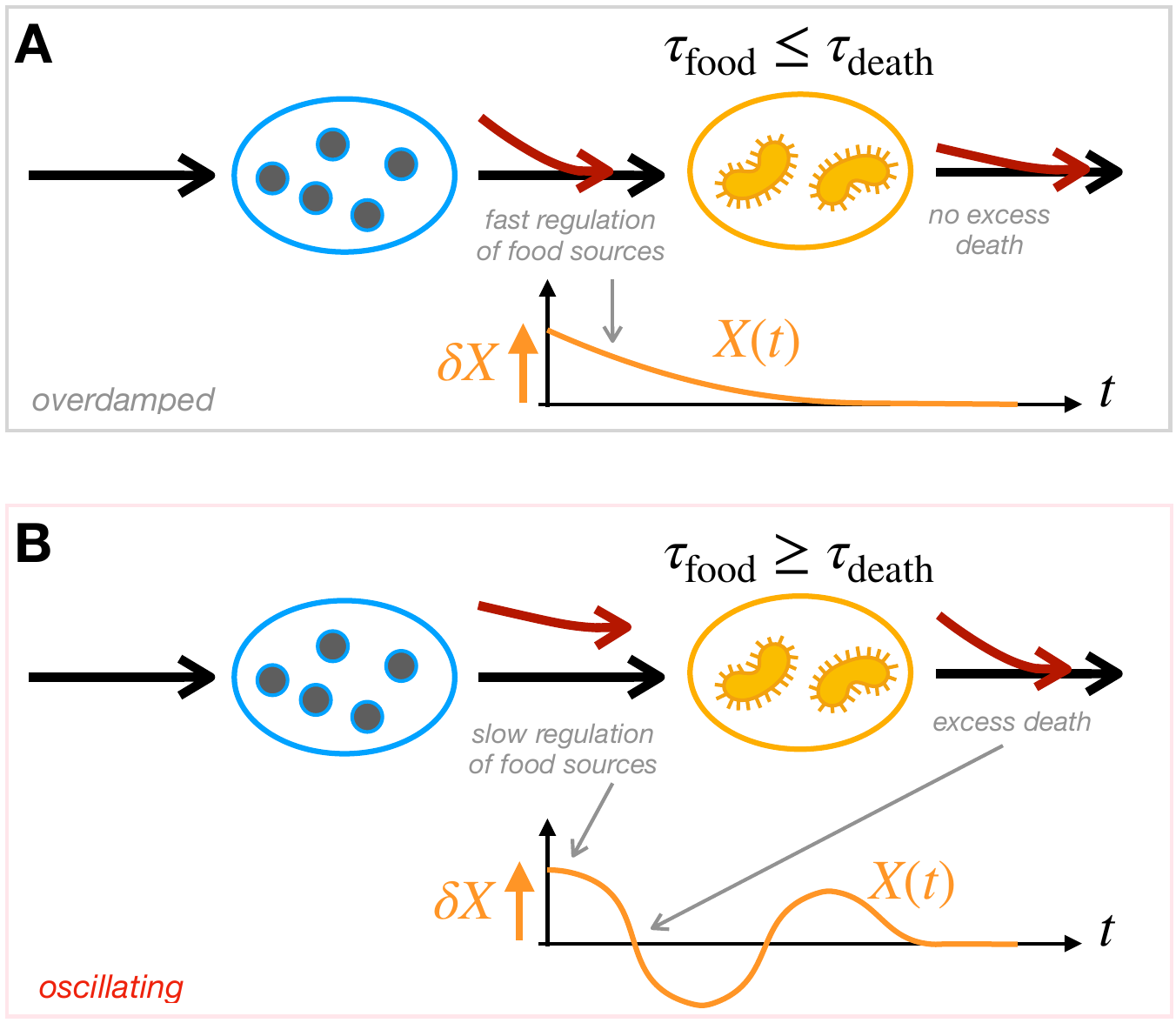}
    \caption{\textbf{Two timescales govern the dynamics of the reference model in Eqs.~(1-2)}. (A) Schematic: in the overdamped scenario, a fast regulation of food sources prevents excess death, resulting in a stable approach of equilibrium. (B) In the oscillating case, the regulation of food sources is slow, resulting in excess death response and creating oscillatory response towards equilibrium.} 
    \label{fig:figure4}
\end{figure}

In practice, the emergence of oscillations via the above mechanism is only possible if the system is slow enough when it responds to changes in substrate concentration, the ``upstream'' phenomenon, compared to the ``downstream'' phenomenon of microbial death, introducing a delay that causes these oscillatory cycles -- see schematic illustration in Fig.~\ref{fig:figure4}. 
This can be translated by a condition on timescales. There are two relevant timescales in the process. When the food concentration changes by a small amount, say $\ccn = \ccn^* + \delta \ccn$, then the timescale that defines the typical time for the microbe population to adapt to this change in food abundance is
\begin{equation}
\tauc = \left( \frac{\partial \frac{d\cb}{dt}}{\partial \cc} \bigg|_{*}\right)^{-1} = 1/ r \mu'(\ccn^*) \cbn^*.
\label{eq:tauc}
\end{equation}
This timescale is not related to an individuals' capability to adapt to relative changes in food abundance, rather, it is a trait associated with the population as a whole, and how fast the population of microbes can increase or decrease to accommodate for such changes in food abundance. It thus depends on population level traits such as reproduction rate and population level response curves to different abundances in food (through \textit{e.g.} Monod kinetics). The second timescale is obtained in a symmetric fashion and is associated with regulation of food supplies by microbes. It is given by 
\begin{equation}
\taud = \left( \frac{\partial \frac{d\cc}{dt}}{\partial \cb} \bigg|_{*} \right)^{-1}= 1/\death
\label{eq:taud}
\end{equation}
and in this case, is equal to the microbial death rate. $\tauc$ and $\taud$ define the two timescales which most control the system response.

These timescales are compared through the parameter $\beta$, since by Eq.~\eqref{eq:beta}
\begin{equation}
    \beta = \frac{r \mu'(\ccn^*)\cbn^*}{\death} \equiv \frac{\taud}{\tauc}
    \label{eq:timescales}
\end{equation}
can indeed be interpreted as a ratio of timescales. The condition for the emergence of oscillations, $\beta \leq 4$, translates on a condition on timescales as
\begin{equation}
    \text{if} \,\, \tauc \gtrsim \taud, \text{oscillations occur.}
    \label{eq:condition}
\end{equation}
When the upstream phenomenon is slow compared to the downstream one, we might expect that the system does not respond fast enough to prevent oscillations, and reciprocally. In this paper, we will show how the condition given in Eq.~\eqref{eq:condition} can be used repeatedly to interpret behavior in various scenarios. 

Finally, we further interpret the stability around the fixed point. Whether through oscillations or directly damped, solutions reach the fixed point with a decay rate that increases with $\beta$: the more responsive the system, the faster the decay, as could have been intuitively expected. In general, the equations are built in a self-regulating way, such that an excess of resources translates in microbial growth which in turn regulates the amount of resources until the system finds a steady state. When $\beta \geq 4$, the system responds quickly, avoiding delays between resource consumption and growth, and is also damped faster. 
The period of the oscillations is given by $1/\Omega \sim 1/\sqrt{\beta}$ and is consistent with this ``time delay'' interpretation.


In the canonical case of Monod kinetics, $\mu(s)~=~s/(1~+~s)$, the fixed point is 
\begin{equation}
    (\cbn^*,\ccn^*) =   \left( \frac{\influx}{\death}, \frac{1}{r/\death - 1} \right), 
    \label{eq:monodstat}
\end{equation}
and the timescales become
\begin{equation}
 \tauc = \frac{\death}{r\influx} \left( 1 - \frac{\death}{r} \right)^{-2}, \, \taud = \frac{1}{\death}.   \label{eq:betaMonod}
\end{equation}
The condition $\beta = \taud/\tauc \leq 4$ for the emergence of oscillations translates into a constraint on $\influx$ and $r$.
In practice, we can always expect the reproduction and growth rate to be larger than the death rate $r \geq \death$, since otherwise a stable population could not be sustained. This simplifies Eq.~\eqref{eq:betaMonod} such that 
\begin{equation}
 \tauc \simeq \frac{\death}{r\influx}, \, \taud = \frac{1}{\death}.   \label{eq:betaMonod2}
\end{equation}
When substrate is flowed in at a higher rate, so when $\influx$ increases, this increases the stable population size, see  Eq.~\eqref{eq:monodstat}. As a result, the population can adapt more quickly to changes in substrate concentration, as indicated via Eq.~\eqref{eq:betaMonod2}. Higher substrate inflow rates can thus prevent oscillations, and increase the rate at which the population converges to the steady state.




\subsection{Link with persistent oscillations: noisy inflow of food}

We now discuss the relevance of studying the emergence of such damped oscillatory solutions around fixed points. In the deterministic case that we studied so far, the oscillations are damped. This is different from sustained population oscillations, which are the source of much fascination in the literature~\cite{fowler2014oscillations,fowler2014starvation,huisman1999biodiversity,khatri2012oscillating}. However, we demonstrate here that the presence of damped oscillations in the deterministic case is correlated with the occurrence of sustained oscillations in response to a noisy excitation. 

To explore the behavior of the system in response to noise, we consider its response to a noisy inflow of food. This is motivated by biophysical considerations whereby sources of food may come in noisy ways due to various external conditions. Note that quite similar response may be found by adding noise on both food sources and bacteria population in a ``statistical physics'' consistent way~\cite{khatri2012oscillating}. However, given noise sources in this context are hardly in equilibrium, there is no reason \textit{a priori} that noise should satisfy fluctuation dissipation. 

We thus explore the behavior of the system near the fixed points in the presence of a noisy inflow of food. Specifically, we modify Eq.~\eqref{eq:framework} to reflect noisy inflow as
\begin{equation}
\begin{split}
    \dtn{\cbn} &= \frac{r}{\death} \mu(\ccn) \cbn -  \cbn  \\
    \dtn{\ccn} &= \frac{\influx(1 + \epsilon\xi(t))}{\death} - \frac{r}{\death} \mu(\ccn) \cbn 
    \label{eq:frameworknoise}
    \end{split}
\end{equation}
where $\xi(t)$ is a nondimensional noise source and where $\epsilon$ is a nondimensional number characterizing the amplitude of the noise. We can choose for instance a Gaussian white noise source of the form $\langle \xi(t) \rangle = 0$ and $\langle \xi(t) \xi(t') \rangle = \tau_n \delta(t-t')$, where $\tau_n$ a typical timescale associated with the noise source. While different authors explore the addition of noise to the system in different ways~\cite{khatri2012oscillating}, here we rather choose to investigate the impact of fluctuating food sources by implementing noise via Eq.~\eqref{eq:frameworknoise}. 

To study this system of equations we study the system around equilibrium, still given by Eq.~\eqref{eq:fixedGeneral}. Then we study the system around equilibrium assuming $\cbn = \cbn^* + \epsilon \chi$ and $\ccn = \ccn^{*} + \epsilon \sigma$. We insert this form in Eq.~\eqref{eq:frameworknoise} and keep only highest order terms in $\epsilon$, such that
\begin{equation}
    \begin{split}
    \dtn{\chi} &=  \beta \sigma,  \\
    \dtn{\sigma} &= \frac{\influx}{\death}\xi - \chi - \beta \sigma.
    \end{split}
\end{equation}
To go further, we introduce the Fourier transform $\tilde{f} = \int \mathrm{d} t f(t) e^{i\omega t }$, and the Power spectral density of a function $I_f(\omega)  = \int \langle f(t) f(0)  \rangle e^{i \omega t}$ is the Fourier transform of its autocorrelation function. With standard calculus one finds the power spectra of both food sources and microbial population as
\begin{equation}
    I_\sigma (\omega) = \frac{\omega^2}{\beta^2} I_{\chi} (\omega) = \frac{\frac{n^2}{\lambda^2} \omega^2 \tau_n}{(\beta - \omega^2)^2 + \omega^2 \beta^2}. 
    \label{eq:snoise}
\end{equation}
By investigating the solutions' denominator, we find it can reach a maximum corresponding to a resonant frequency $\omega_c  = \sqrt{2\beta - \beta^2}$, which only exists when $\beta \leq 2$. Under that condition, there is thus a resonance between the inflow noise and the systems' dynamics that drives more persistent oscillations. Remarkably $\beta \leq 2$ is also the condition in the deterministic case to witness significant oscillations in the decay towards equilibrium. 

\begin{figure}[h!]
    \centering
    \includegraphics[width=0.99\linewidth]{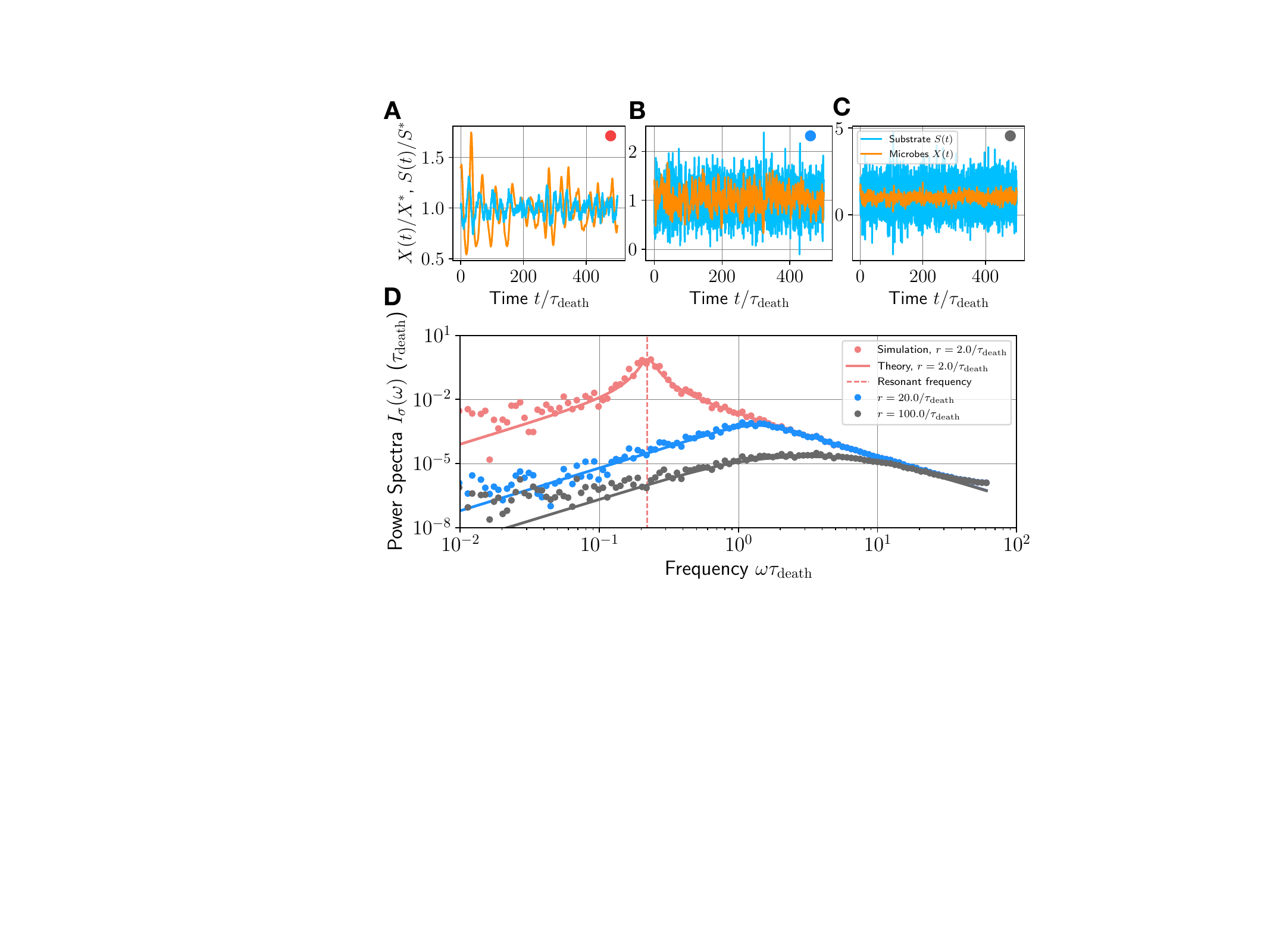}
    \caption{\textbf{Emergence of persistent oscillations in the presence of noise}. (A, B, C) Typical solutions in the presence of $10\%$ noise source on food inflow for the same parameter values as in Fig.~\ref{fig:figure4} (C, D, E) indicated by the colored dots. (D) Corresponding calculated noise spectra of the food source, with lines indicating theory curves from Eq.~\eqref{eq:snoise}. For all this figure $n = 0.1\ /\taud$. } 
    \label{fig:figure4a}
\end{figure}

In Fig.~\ref{fig:figure4a} we illustrate these findings by looking at the response of the system to a small noisy food inflow, around steady states. We investigate the microbial and food resource dynamics in a case where $\beta \leq 2$, corresponding to the ``oscillating'' domain in the deterministic case (Fig.~\ref{fig:figure4a}-A); $2 \leq \beta \leq 4$, the ``damped'' domain (Fig.~\ref{fig:figure4a}-B); and $4 \leq \beta$, the ``overdamped'' domain (Fig.~\ref{fig:figure4a}-C). We find that oscillations develop around the steady state only in case A, consistently with our predictions. This is correlated with the presence of a peak in the power spectral density of the food source (red in Fig.~\ref{fig:figure4a}-D), where the location of the peak corresponds to the resonant frequency $\omega_c$. 

Damped oscillatory dynamics in the deterministic case are thus correlated with persistent oscillatory dynamics in noisy cases. This would also be true for periodically driven cases, which would get resonant oscillatory response at a frequency $\omega = \omega_c$. 
For all these reasons, it is thus sensible to rationalize how different biophysical factors modify the parameter space where damped oscillations occur in the deterministic case, as a proxi for persistent oscillatory response in noisy or driven cases. 
We thus explore various modifications of the model to include common yet more complex bio-physical features, specifically: recycling of food resources via microbial necromass in Sec.~\ref{sec:recycling}; dependence on multiple food resources in Sec.~\ref{sec:foods}; impact of the microbes' environment especially of further species in Sec.~\ref{sec:environment}. We will show that, mathematically, modifications can easily be captured within the same framework, especially condition Eq.~\eqref{eq:condition}, and conclusions on oscillatory behavior drawn without requiring more complex calculations. This will allow us to identify physical phenomena that prevent or facilitate oscillatory behavior. 

\section{Recycling via microbial necromass}
\label{sec:recycling}

We first explore behavior when part of the microbial necromass -- or deceased microbes -- is converted back into food sources, see for example Refs.~\cite{sanchez2013analytical,fowler2014oscillations,camenzind2023formation}. 

\subsection{Simple model of recycling}

 To account for such ``recycling'' we can modify Eq.~\eqref{eq:model1substrate} as
\begin{equation}
        \dt{\cc} = \influx \cco -  \gamma  r \mu(\cc)\cb + \alpha \gamma \death \cb \label{eq:model2substrate}
\end{equation}
where $\alpha < 1$ is a nondimensional number characterizing the conversion of necromass into reuseable food sources. This model typically supposes that there is no delay in the conversion from dying microbes to useable food sources, as several authors suppose~\cite{fowler2014oscillations,sanchez2013analytical}. However, because microbial decay takes time, it could be more relevant to model microbial necromass as a separate entity, which we will do in the following section. We now non-dimensionalize variables as $\tau = \death_{\alpha} t$, and introduce an effective decreased death rate $\death_{\alpha} = \death (1-\alpha)$, and an effective function $\mu_{\alpha}(\ccn) =  \mu(\ccn) - \alpha \death_{\alpha}/(1-\alpha)r $ such that we obtain a system of equations exactly similar to Eq.~\eqref{eq:framework}, as
\begin{equation}
\begin{split}
    \dtn{\cbn} &= \frac{r}{\death_\alpha} \mu_{\alpha}(\ccn) \cbn -  \cbn  \\
    \dtn{\ccn} &= \frac{\influx}{\death_\alpha} - \frac{r}{\death_\alpha} \mu_{\alpha}(\ccn) \cbn.
    \end{split}
    \label{eq:recycling}
\end{equation}
Without any assumption for now on the shape of $\mu(s)$, the fixed point can be obtained from Eq.~\eqref{eq:fixedGeneral}, then simplified to
\begin{equation}
    (\cbn^*_{\mathrm{\alpha}},\ccn^*_{\mathrm{\alpha}}) = \left( \frac{\influx}{\death_{\alpha}}, \mu_{\alpha}^{-1} \left( \frac{\death_{\alpha}}{r}\right) \right) = \left(\frac{\cbn^*}{1-\alpha},\ccn^*\right)
    \label{eq:fixedrecycling}
\end{equation}
and we remark that this fixed point only slightly change compared to the reference case.
Recycling contributes to increase the steady-state microbial population, as one would expect, by increasing the effective abundance of substrates, through both substrate inflow and microbial necromass. The steady state substrate concentration does not change. 
The timescales become
\begin{equation}
\begin{split}
    &\tauc^{(\alpha)} = \left(r \mu_{\alpha}'(\ccn_{\alpha}^*)\cbn^*_\alpha \right)^{-1} = 
    \tauc(1-\alpha) \, \\
    &\taud^{(\alpha)} = \frac{1}{\death_{\alpha}} = \frac{\taud}{1-\alpha}. 
\end{split}
\end{equation}
The timescale to adapt to changes in substrate concentration decreases due to recycling, in part because the sustained steady state population is larger and in part because the system is quicker to adapt to changes thanks to the recycling process. The effective death rate is also reduced, because part of microbial death serves to feed back positively on microbial growth. 

The parameter range over which oscillations occur in phase space is thus reduced. In fact $\beta_{\alpha} = \taud^{(\alpha)}/\tauc^{(\alpha)} = \beta/(1-\alpha)^2$, and the condition $\beta_{\alpha} \leq 4$ translates into a condition $\beta \leq 4(1-\alpha)^2$: the range over which oscillations occur is narrower. In addition, the oscillations decay with a damping rate as $\exp( - \beta_{\alpha} \tau) = \exp( - \beta \lambda t/(1-\alpha))$ and thus adaptation is also faster with recycling. Finally, the oscillations occur with modified frequency, according to Eq.~\eqref{eq:frequency}, $\Omega^{(\alpha)}/\death_\alpha = \sqrt{4 \beta_\alpha - \beta_\alpha^2}/2$ which simplifies to
\begin{equation}
    \frac{\Omega^{(\alpha)}}{\lambda} = \sqrt{4 \beta - \frac{\beta^2}{2(1-\alpha)^2}}
\end{equation} which is always smaller than $\Omega$ in Eq.~\eqref{eq:frequency} in the reference case. The oscillations are thus slower than in the reference case.
One can therefore consider the system is more ``stable'' in the presence of recycling. 

\begin{figure}
    \centering
    \includegraphics[width=0.99\linewidth]{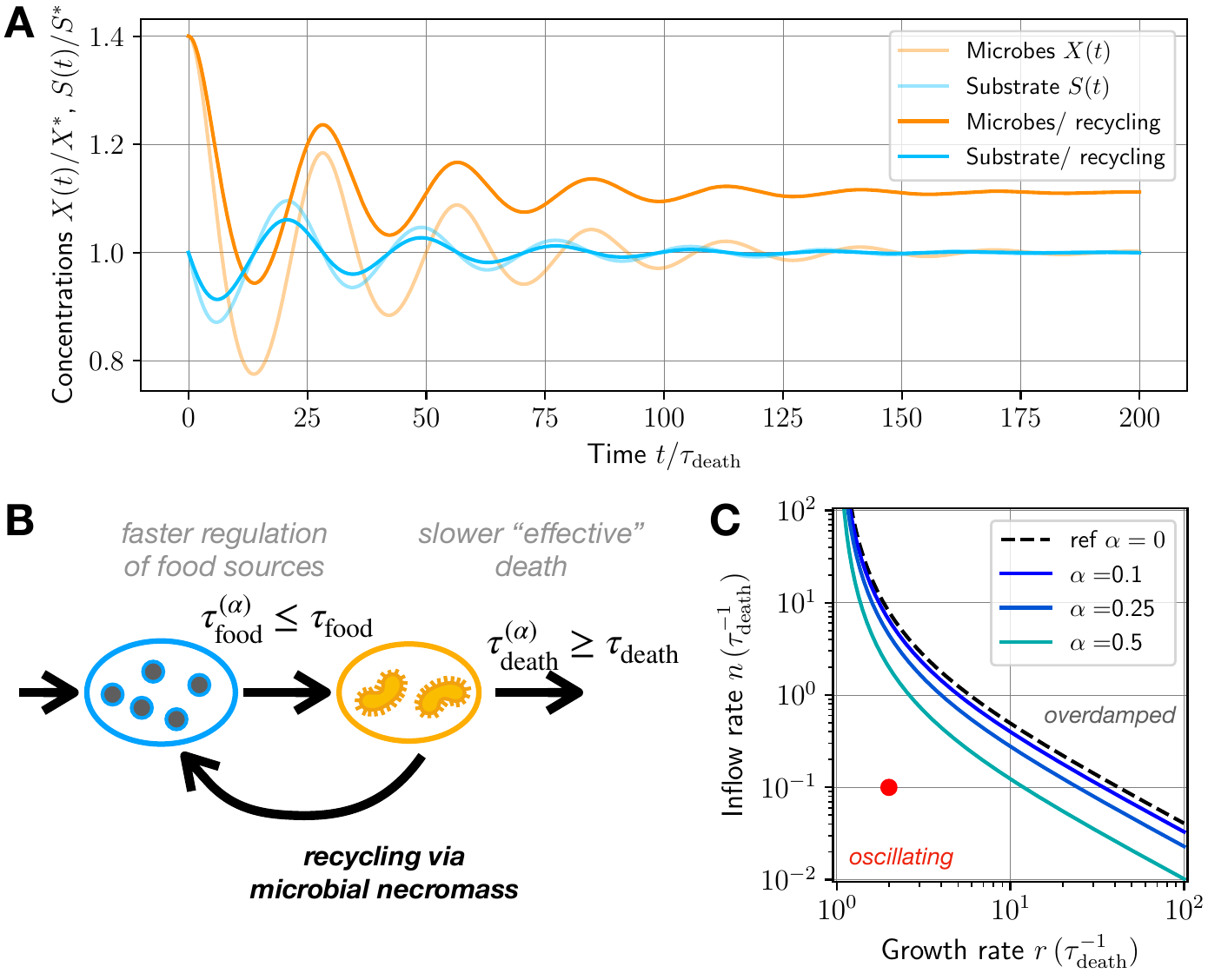}
    \caption{\textbf{Impact of recycling on oscillatory solutions.} (A) Typical solutions with recycling, comparing to the reference case (lighter color) of Fig.~\ref{fig:figure3}-C without recycling. Parameters are the same as in Fig.~\ref{fig:figure3}-C and with $\alpha = 0.2$. (B) Schematic impact of recycling: by providing new food sources, recycling speeds up food source regulation. (C) Phase diagram similar as in Fig.~\ref{fig:figure3}-A, with separatrix lines corresponding to $\beta_{\alpha} = 4$. The red dot corresponds to the point in phase space where the simulation in (A) was done.}
    \label{fig:figure5}
\end{figure}

A typical example of an oscillatory solution in the recycling case, with $\alpha = 0.2$ is shown in Fig.~\ref{fig:figure5}-A (opaque lines). In this case we assumed Monod kinetics for simulation purposes. The reference case where $\alpha = 0$ is also illustrated with transparent lines, and appears to be damped much more slowly than the case with recycling. The change in oscillatory frequency is not obvious because in the parameter range explored, it is small. The most striking feature is the change in steady state microbial population, which is larger in the case with recycling. Fig.~\ref{fig:figure5}-C presents lines on the parameter phase diagram, differentiating regions with oscillating or overdamped solutions relative to the reference case, clearly illustrating the narrowing of the oscillatory region with recycling. The red dot represents the point in phase space corresponding to Fig.~\ref{fig:figure5}-A. 

Why does recycling lead to longer oscillations but faster decay rates? Consider again for interpretation a hypothetical scenario where biomass is in slight excess compared to the fixed point, but substrate concentration is at steady state $\cc^*$. The rate of substrate consumption thus exceeds the inflow rate but not as much because recycling helps to sustain the current over-populated state: the substrate concentration decreases more slowly than in the no-recycling case -- see Fig.~\ref{fig:figure5}-B. Similar reasoning applies to the next steps of the cycle. This essentially demonstrates that the oscillations occur slower than without recycling because the imbalance between substrate/microbial concentrations is slower to establish and invert. However, because the system responds slower, adaptation to the steady state is faster. 

\subsection{Delayed recycling model}

\begin{figure*}
    \centering
    \includegraphics[width=0.99\linewidth]{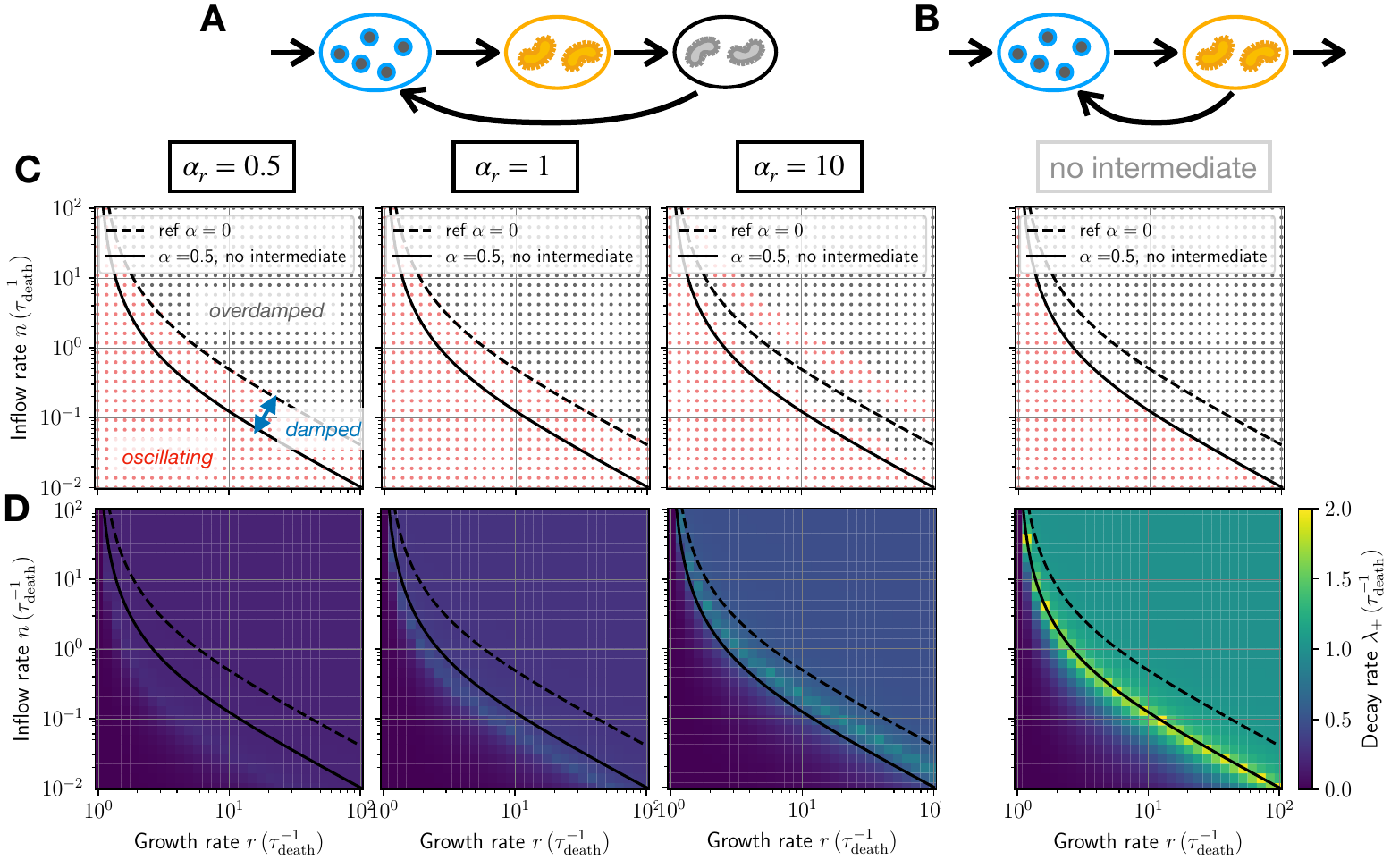}
    \caption{\textbf{Delayed necromass recycling is correlated with slowed dynamics. } (A) Schematic of necromass recycling with a delayed usability of necrommas and (B) if immediate necromass usability. (C) Numerically obtained phase diagrams for different cases in (A-B) keeping $\alpha = 0.5$ but variable $\alpha_r$. The red dots correspond to cases where the jocabian matrix near the fixed point has eigenvalues with complex components, and the black dots where it has only real components. (D) Numerically obtained decay rates, which is the absolute value of the largest real part of the eigenvalues of $J$, for the same cases as in (C). }
    \label{fig:figure5a}
\end{figure*}

A model which could reflect more the time it takes for microbial necromass to be useable as a food source could explicitly model the presence of another component, the microbial necromass $M$ (see Fig.~\ref{fig:figure5a}-A and B), which is converted into useable food sources at a given rate $\lambda_r$. In the nondimensional scales, the necromass is $m = M/\gamma K_S$ and the nondimensional equations satisfied by the system are now
\begin{equation}
    \begin{split}
        \frac{d\cbn}{dt} &= \frac{r}{\death} \mu(\ccn) \cbn - \cbn \\
        \frac{dm}{dt} &= \alpha_r\left(\alpha \cbn -  m\right) \\
        \frac{d\ccn}{dt} &= \frac{\influx}{\death} - \frac{r}{\death} \mu(\ccn) \cbn + m.
        \end{split}
\end{equation}
where $\alpha_r = \lambda_r / \death$ characterizes the ratio between the necromass recycling rate $\lambda_r$ and microbial death $\death$. Note that the limit $\alpha_r = \infty$ corresponds to the set of Eq.~\eqref{eq:recycling}. 
The nontrivial fixed point in this delayed recycling case is
\begin{equation}
    (\cbn^*_\alpha,m^*_\alpha,\ccn^*_\alpha) = (\frac{\cbn^*}{1 - \alpha},\cbn^* \alpha,\ccn^*) 
\end{equation}
which is similar to Eq.~\eqref{eq:fixedrecycling}. The dynamical behavior near the fixed point can be explored via the Jacobian matrix, 
\begin{equation}
    J = \begin{pmatrix}
        0 & 0 & \beta \\
        \alpha_r \alpha & - \alpha_r & 0 \\
        -1 & 1 & -\beta
    \end{pmatrix}.
\end{equation}
Unfortunately, the behavior of the system around equilibrium is not amenable to analytical calculations but can be solved by looking numerically for the eigenvalues of $J$. 

\begin{figure*}
    \centering
    \includegraphics[width=0.99\linewidth]{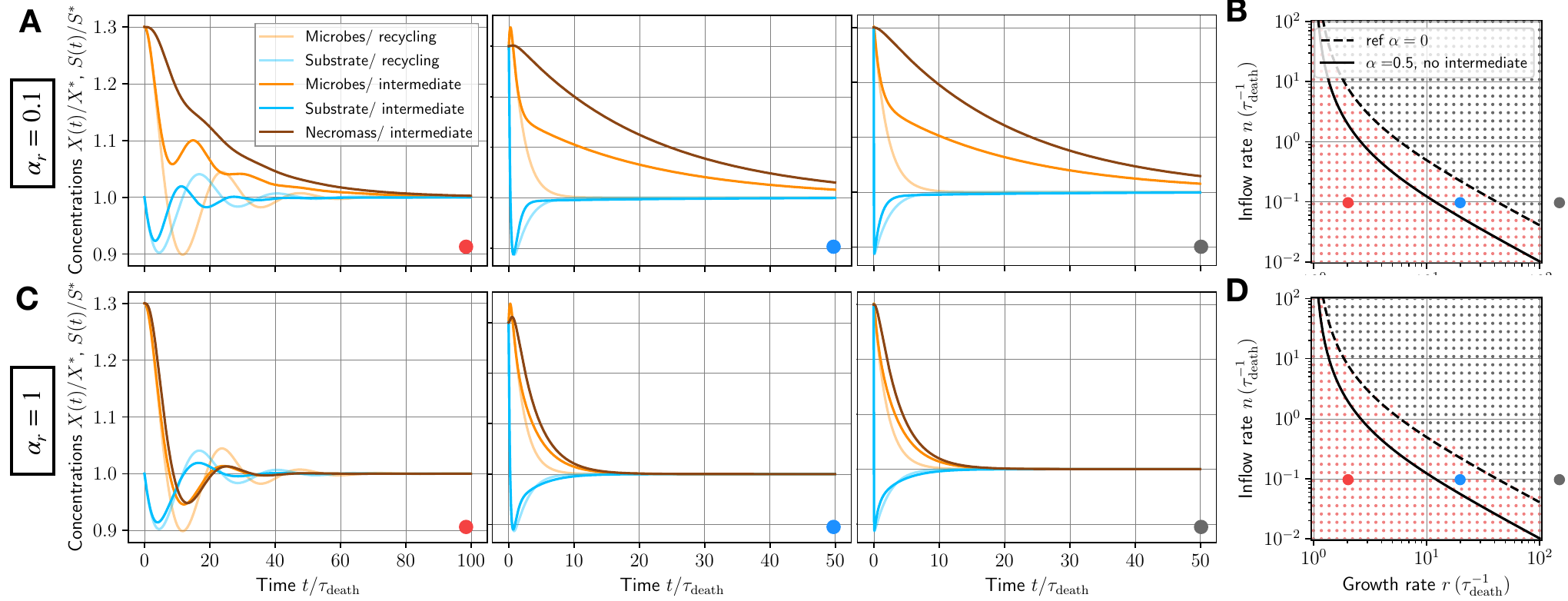}
    \caption{\textbf{Typical response of the system with intermediate necromass and without.} (A-C) representative numerical solutions obtained for the case without intermediate (transparent lines) and with intermediate (opaque lines), for 2 different values of $\alpha_r$, and 3 different parameter values illustrated with colored points, and reported in the phase diagrams (B,D). Red: $r = 2 \ /\taud$, blue: $r = 20 \ /\taud$, gray: $r = 200 \ /\taud$ and $n = 0.1 / \taud$ in all cases. The phase diagrams (B-D) are the same as in Fig.~\ref{fig:figure5a}.}
    \label{fig:figure5b}
\end{figure*}

We thus build numerical phase diagrams in Fig.~\ref{fig:figure5b}. Typically, the phase diagrams obtained are very similar to the case with no intermediate microbial necromass, meaning such delayed response does not modify significantly the range of dynamic types in phase space. We will explore a bit more detailed differences with Fig.~\ref{fig:figure5b}. The most striking difference is rather obtained by investigating the decay rate towards steady state. Here, even with $\alpha_r = 1$, the decay rate is orders of magnitude smaller compared to the case with no intermediates. This means that the dominant effect of the delayed response is to slow down the system's response. 

Further insights may be obtained by exploring typical response curves of the system, reported in Fig.~\ref{fig:figure5b}. The slowed down system response is generically apparent, especially in the $\alpha_r$ case, where full lines representing the different components in the case with the intermediate necromass relax much slower towards equilibrium than transparent lines in the case without intermediate necromass. These representative cases also allow us to explore the intermediate regimes of parameters (blue dots) where an oddity occurs. In this blue region, the analysis of $J$ (with intermediate necromass) predicts an oscillatory response, however this occurs in the overdamped region of the phase space without intermediate necromass. In that case, we see in Fig.~\ref{fig:figure5b}-A and C, blue dot case, that a minute oscillation occurs before the system is slowly damped, essentially, the oscillations don't have time to establish. 

Therefore, the range of parameter space where oscillations occur or not is independent of the delayed necromass cycling. In fact, in the delayed case, we can calculate $\tauc^{(\alpha)}$ and $\taud^{(\alpha)}$ and find they do not change compared to the case without intermediate. This explains physically why the range of oscillatory or not dynamics should also not change significantly.

\section{Diversity of food supplies}
\label{sec:foods}

Even if a microbial species is the single species in a given environment, it relies for survival on different food sources. Some of these food sources are substitutable or complementary, meaning that a species can grow interchangeably on one or the other while some are essential or necessary, meaning that both resources are needed for growth~\cite{tilman1982resource}. We explore the response of our reference system to different scenarios where multiple food sources are at play.

\subsection{Complementary food sources}

\begin{figure}
    \centering
    \includegraphics[width=0.99\linewidth]{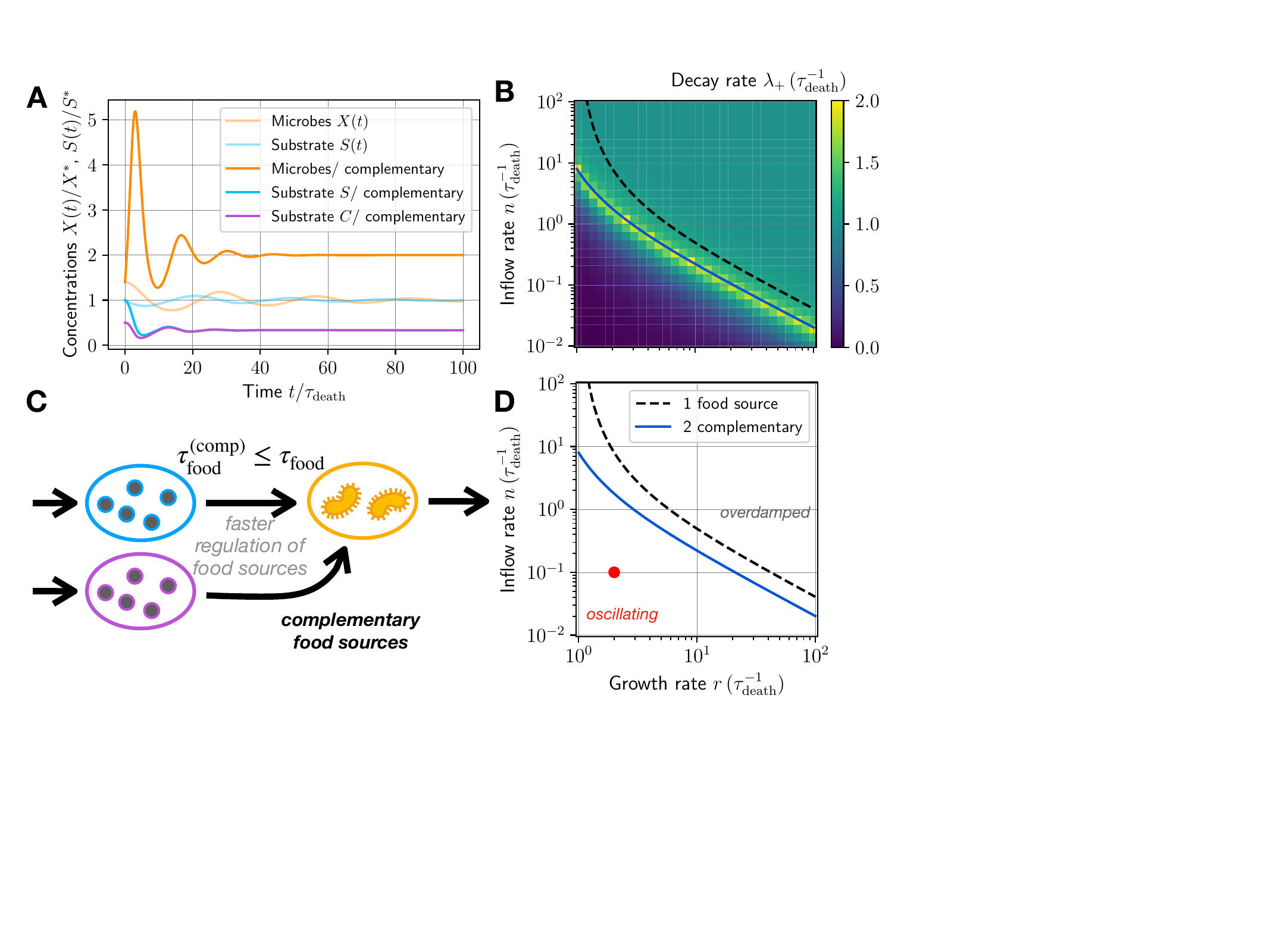}
    \caption{\textbf{Impact of adding a complementary food source on oscillatory solutions.} (A) Typical solution with complementary food sources, comparing to the reference case (lighter color) of Fig.~\ref{fig:figure3}-C. Parameters are the same as in Fig.~\ref{fig:figure3}-C and the same for the two complementary food sources. (B) Decay rate in phase space for the case of two complementary food sources. (C) Schematic impact of complementary food sources: by providing more and directly edible food sources, food source regulation is sped up. (D) Phase diagram with the separatrix lines between the oscillating and overdamped regimes, $\beta_{\rm comp} = 4$ ; with the red dot showing the parameters used in (A). The separatrix lines are the same in (B).}
    \label{fig:figure6}
\end{figure}

Some food sources can be \textit{complementary}, meaning an organism can rely on one or the other for growth. This is typically the case for carbon resources and bacteria growth, where multiple carbon sources, or differnt kinds of sugars, may be used to grow~\cite{bajic2020ecology,fowler2014oscillations,fowler2014starvation,tilman1982resource}. This can also be the case for different kinds of nitrate resources~\cite{gowda2022genomic}. To account for 2 complementary food sources we write the following set of equations 
\begin{equation} \begin{split}
    \dt{\cb} &= r \mu_{\cc}(\cc) \cb + r \mu_{\ccd}(\ccd) \cb - \death \, \cb \label{eq:modelcomplementary} \\
    \dt{\cc} &= \influx \cco -  \gamma  r \mu_\cc(\cc)\cb  \\
    \dt{\ccd} &= \influx \ccdo -  \gamma  r \mu_{\ccd}(\ccd)\cb  
\end{split}\end{equation} 
where $C$ refers to the concentration of the second food source (see also schematic in Fig.~\ref{fig:figure6}-B). For simplicity, we considered, for now, that $r, \influx,$ and $\gamma$ are similar for both food sources. In this model, we took a form where contributions are essentially summed up, similarly as in Ref.~\cite{fowler2014oscillations,fowler2014starvation,gowda2022genomic}, however one could consider different analytical approaches~\cite{haas1994unified}. 

Nondimensionalizing $c = C/K_C$ and $\ccn = \cc/K_S$ it is obvious that $\ccn$ and $c$ satisfy similar dynamics. We assume further that $\mu_Z(Z/K_Z)$ does not depend on the specific substrate $Z$. This is not such a restrictive assumption since one notices this is the case for Monod kinetics, since then $\mu_Z(Z) = \frac{Z/K_Z}{Z/K_Z + 1}$. We can then obtain a simplified dynamical system as
\begin{equation} \begin{split}
    \dtn{\cbn} &=  \frac{r}{\death}  (\mu(\ccn) + \mu(c) ) \cbn -  \cbn  \\
    \dtn{\ccn} &= \frac{\influx}{\death} - \frac{r}{\death} \mu(\ccn) \cbn \\
    \dtn{c} &= \frac{\influx}{\death} - \frac{r}{\death} \mu(c) \cbn.
    \label{eq:complementary}
\end{split}\end{equation} 
The non trivial fixed point of the system is given by
\begin{equation}
    (\cbn^*_{\rm comp},\ccn^*_{\rm comp},c^*_{\rm comp}) = \left( 2 \cbn^*, \mu^{-1}\left( \frac{\lambda}{2r}\right),\mu^{-1}\left( \frac{\lambda}{2r}\right) \right). 
\end{equation}
The population sustained by this system is doubled since the substrate influx is doubled. In the case of Monod kinetics, we would find $\ccn^*_{\rm comp} = c^*_{\rm comp} = 1/ (2 r/\death - 1)$ , the steady-state substrate concentration for each compound than that in the case of a single food source because the steady state population is larger. 

\begin{figure*}
    \centering
    \includegraphics[width=0.99\linewidth]{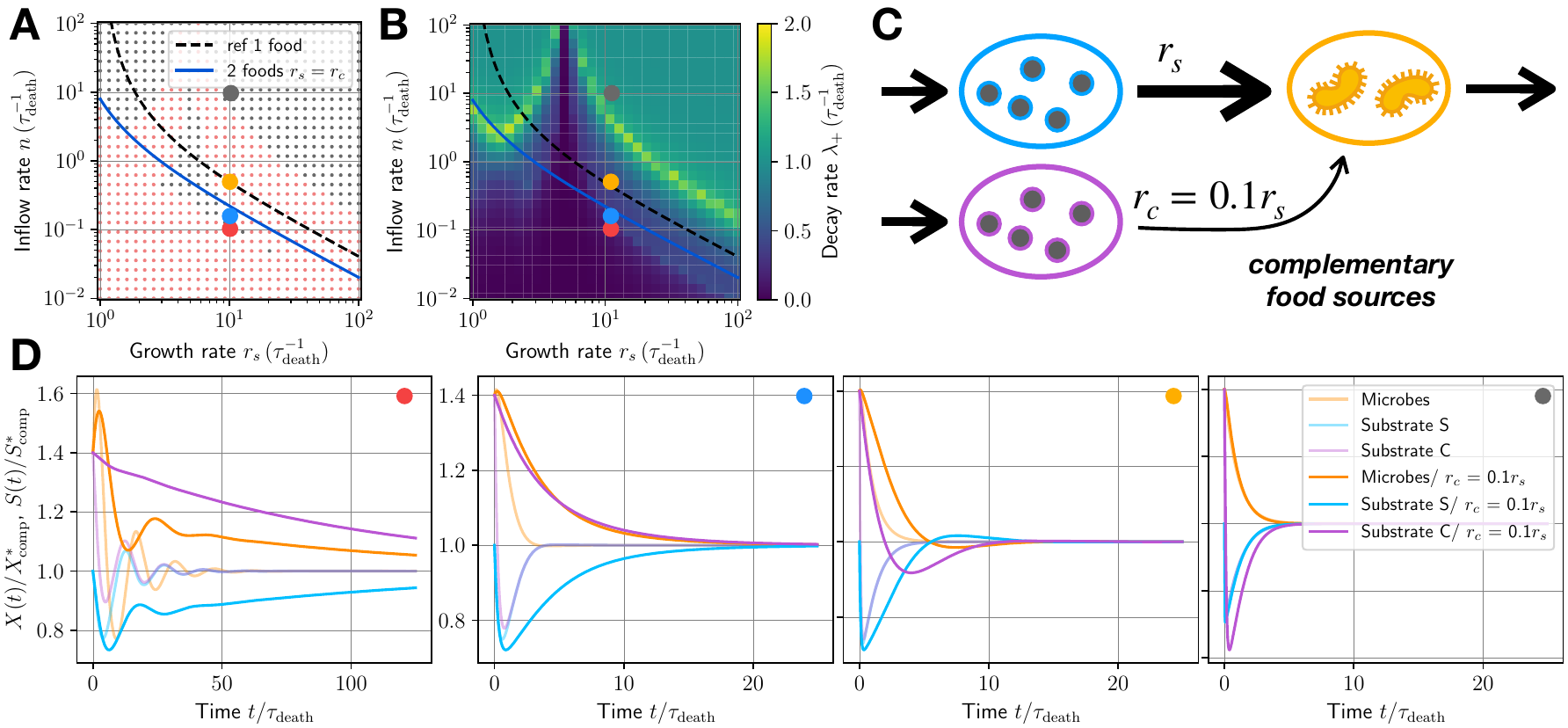}
    \caption{\textbf{Diverse complementary food sources.} (A) Numerically obtained phase diagram and (B) decay rate in the case of two complementary food sources where one food source is preferred over the over. The numerical procedure is similar as the one used to obtain Fig.~\ref{fig:figure5a}. For both (A-B), the growth rate taken for plots is $r_s$ and we keep a fixed $r_c = 0.1 r_s$, as sketched in (C). (D) Representative numerical solutions around the steady state for the different points in phase space illustrated by the colored dots in (A-B).  Transparent lines in each plot correspond to the reference case where $r_c = r_s$. In these plots, $r_s = 10 \ /\taud$ and $n = [0.1, 0.15, 0.5, 10] \ /\taud$. }
    \label{fig:figure6b}
\end{figure*}

Following the framework derived above, we can find the Jacobi matrix and that damped oscillations towards the fixed point exist if $\beta_{\rm comp} = \taud^{\rm (comp)}/\tauc^{(\rm comp)} \leq 4$ (see Appendix~\ref{app:complementary}).
The timescale for adaptation to changes in substrate concentration becomes
\begin{equation}
\begin{split}
    \tauc^{(\rm comp)} = &\left( r \mu'(\ccn_{\rm comp}^*)\cbn^*_{\rm comp} \right)^{-1} = \tauc f\left(\frac{r}{\death}\right) \\
     & \text{where} \,\,f(y) = \frac{2 \left( y -1 \right)^2}{\left(2 y -1 \right)^2}
\end{split}
\end{equation}
where the last equality is obtained assuing Monod kinetics. 
while the timescale related to death does not change $\taud^{\rm (comp)} = \taud$. 
One can easily show that for $y = r/\death \geq 1$, then $f(y) \leq 1/2$. The time to adapt to substrate concentration changes is thus shorter with multiple food sources. This is due in part to the fact that the total sustained population is larger but also the two food sources act as complementary reservoirs when one food source is missing, enabling faster recovery. The reactivity parameter in that context can be reexpressed relative to our general case as $\beta_{\rm comp} = \beta/f(r/\lambda)$, 
such that $\beta_{\rm comp} \geq 2 \beta$. The oscillations occur if $\beta_{\rm comp} \leq 4$ which translates into $\beta \leq 2$, which narrows down the range over which oscillations occur, stabilizing the system. The damping factor is given by $\beta_{\rm comp}$ and hence is always larger than $\beta$. For illustrative purpose we again represent these effects in Fig.~\ref{fig:figure6}. Complementary food sources thus tend to reduce the range of paramters where oscillations arise. 

In more realistic cases, one food source can be preferred over the other for growth, as in the case of \textit{E. Coli} and different sugars~\cite{bajic2020ecology}.
If one food is preferred over the other, then this can be accounted for by choosing growth rates $r_s$ and $r_c$ that are different for each of the food sources $S$ and $C$ respectively in Eq.~\eqref{eq:modelcomplementary}. While the system is then not amenable to analytics, one can still obtain numerical insights on the phase space. We present the phase diagram and decay rates in Fig.~\ref{fig:figure6b}-A and B for a representative case where a given food source (c) is less favorable over another (s), by taking $r_c = 0.1 r_s$ (see Fig.~\ref{fig:figure6b}-C). We also present representative solutions for different points in the phase space in Fig.~\ref{fig:figure6b}-D, presenting with transparent lines the reference case where $r_c = r_s$. 

The predominant feature in a case where one food source is less favored for growth is that fluctuations away from equilibrium are suppressed much more slowly. Since one food source is typically eaten up less quickly, then it simply takes more time for that food source to relax towards equilibrium. The other noticeable feature is regions of phase space where additional oscillatory behavior arises (see the orange dot and associated 3rd plot in Fig.~\ref{fig:figure6b}-D). There, the preferred food source is consumed quickly at first, leaving only the less favored food around. Yet the population does not subsist well on that less favored food source, creating a momentary drop in the population level. This allows preferred food sources to be replenished, and the cycle goes on. This now can be understood in terms of an interplay of timescales. Since $r_c \leq r_s$, then the time to adapt to changes in food abundance for the different types of food satisfies $\tauc^{\mathrm{(comp,c)}} \geq  \tauc^{\mathrm{(comp,c)}}$, and thus there exists a range of parameter space where 
\begin{equation*}
    \tauc^{\mathrm{(comp,c)}} \gtrsim \taud \gtrsim \tauc^{\mathrm{(comp,s)}}
\end{equation*}
and where we can expect some form of oscillatory behavior around steady state.

\subsection{Necessary food sources}

Other groups of food sources are not substitutable between one another, rather, an organism requires multiple food sources for growth. For instance, to transform carbon sources into energy for growth, many microbes require oxygen at the same time~\cite{fowler2014oscillations,chapelle2000ground}. We explore now this case where 2 food sources are essential or necessary for growth. 


When 2 food sources are necessary, we can write the following set of equations
\begin{equation}
\begin{split}
    \dt{\cb} &= r \mu_{\cc}(\cc) \mu_{\ccd}(\ccd)\cb - \death \, \cb \label{eq:modelcomplementary} \\
    \dt{\cc} &= \influx \cco -  \gamma  r \mu_{\cc}(\cc) \mu_{\ccd}(\ccd)\cb  \\
    \dt{\ccd} &= \influx \ccdo -  \gamma r \mu_{\cc}(\cc) \mu_{\ccd}(\ccd)\cb  
\end{split}\end{equation} 
where for simplicity we considered again $\influx,$ and $\gamma$ to be similar for both substrates. The shape of the dynamics that we chose for necessary foods is close to that in \cite{fowler2014oscillations,chapelle2000ground,haas1994unified} but note that some authors prefer to use the $\min(\mu_{\cc}(\cc),\mu_{C}(C))$ according to Liebig's law of the minimum~\cite{baer2006multiple}.
Again, nondimensionalizing $c = C/K_C$ and $\ccn = \cc/K_S$, and assuming the affinity for both food sources is independent of the food source if variables are rescaled in that way, then we obtain the dynamics
\begin{equation}
\begin{split}
    \dtn{\cbn} &=  \frac{r}{\death}  \mu(\ccn) \mu(c)  \cbn -  \cbn  \\
    \dtn{\ccn} &= \frac{\influx}{\death} - \frac{r}{\death} \mu(\ccn) \mu(c)\cbn \\
    \dtn{c} &= \frac{\influx}{\death} - \frac{r}{\death} \mu(\ccn) \mu(c) \cbn.
    \label{eq:modelcomplementarysimplified}
\end{split}\end{equation} 
The non trivial steady state of the system is given by
\begin{equation}
    (\cbn^*_{\rm nec},\ccn^*_{\rm nec},c^*_{\rm nec}) = \left( \cbn^*, \mu^{-2} \left( \frac{\death}{r}\right),\mu^{-2} \left( \frac{\death}{r}\right)\right). 
\end{equation}
The steady-state population sustained by this system is the same as for the reference case. 
In the case of Monod kinetics, we can obtain an expression for $\ccn^*_{\rm nec} =c^*_{\rm nec} = 1/(\sqrt{r/\death} - 1)$.
It also appears that the steady-state substrate concentration for each compound is higher than in the reference case, which is a result of the population being maintained by both food sources at once. 

Following the general framework, we can find the Jacobi matrix and that damped oscillations towards the fixed point exist if $\beta_{\rm nec} = \taud^{\rm (nec)}/\tauc^{(\rm nec)} \leq 2$ (see Appendix~\ref{app:nec}).
The timescale for adaptation to changes in substrate concentration becomes
\begin{equation}
\begin{split}
    \tauc^{(\rm nec)} &= \left( r \mu'(\ccn_{\rm nec}^*)\mu(c_{\rm nec}^*)\cbn^*_{\rm nec} \right)^{-1} = \tauc g\left(\frac{r}{\death}\right) \\
     & \text{where} \,\, g(y) = \frac{\left( y -1 \right)^2}{\sqrt{y} \left(\sqrt{y} -1 \right)^2}
\end{split}
\end{equation}
while the timescale for death does not change. 
One can show that for any value of $y$ then $g(y) \geq 4$. The time to adapt to changes in substrate concentration is thus longer than that in the reference case with multiple necessary food sources. The reactivity parameter can also be written in terms of the general case as $ \beta_{\rm nec} = \beta/g(r/\lambda)$, it is thus always larger than in the reference case.
Oscillations occur when $\beta_{\rm nec} \leq 2$ which translates into $\beta \leq 4 g(r/\lambda)$. Since $g(y) \geq 4$, the range over which oscillations occur is necessarily broader than in the reference case. The damping factor is given by $\beta_{\rm nec}$ and hence always smaller than $\beta$, such that perturbations away from equilibrium take longer to relax. This means necessary food sources tend to increase the range of parameter space where oscillations occur.

\begin{figure}[h!]
    \centering
    \includegraphics[width=0.99\linewidth]{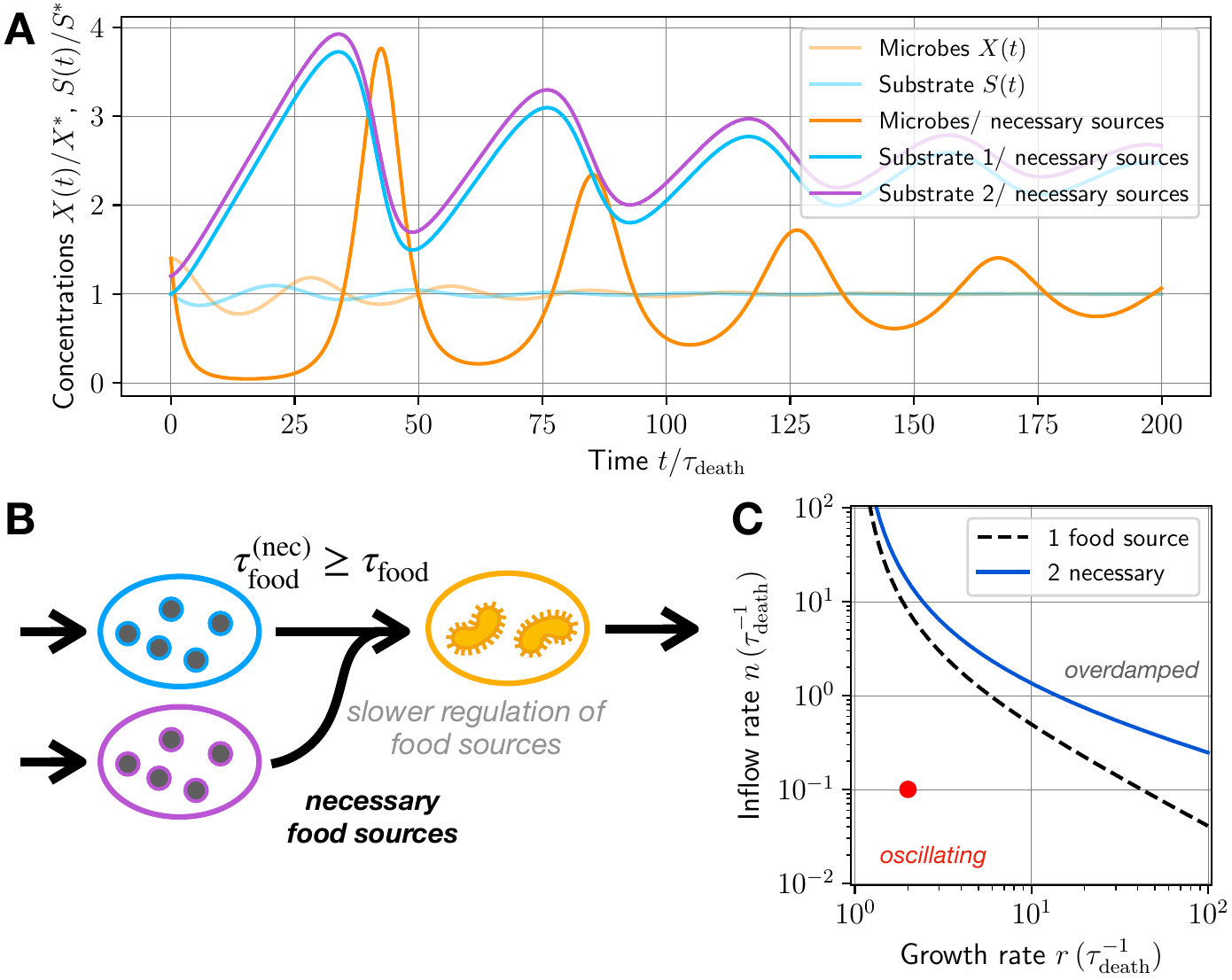}
    \caption{\textbf{Impact of adding a necessary food source on oscillatory solutions.} (A) Typical solution with necessary food sources, comparing to the reference case (lighter color) of Fig.~\ref{fig:figure3}-C. Parameters are the same as in Fig.~\ref{fig:figure3}-C and the same for the two necessary food sources. (B) Schematic impact of necessary food sources: by restricting conditions for growth on two necessary food sources, food source regulation is slowed down. (C) Phase diagram with the separatrix lines between the oscillating and overdamped regimes, $\beta_{\rm nec} = 2$; with the red dot showing the parameters used in (A).}
    \label{fig:figure7}
\end{figure}

This behavior is illustrated by Fig.~\ref{fig:figure7}. In Fig.~\ref{fig:figure7}-A we present a typical oscillatory solution to the dynamics in the ``two necessary food sources'' case, where the parameters are the same as for the oscillatory solution in the reference case in Fig.~\ref{fig:figure3}-C. Clearly, the system undergoes drastic oscillations, where the microbial concentration approaches zero. The mechanism is illustrated in Fig.~\ref{fig:figure7}-B, where the face that two food sources are necessary makes for a slower regulation of food sources upon changes of microbial population levels.

\section{Environmental factors}
\label{sec:environment}

We finally explore the role of several environmental factors on this oscillatory behavior, namely of (A) multiple species; (B) and outflow rate of resources; and (C) consequences on waste production. 

\subsection{Multi-species environment}

\begin{figure}
    \centering
    \includegraphics[width=0.99\linewidth]{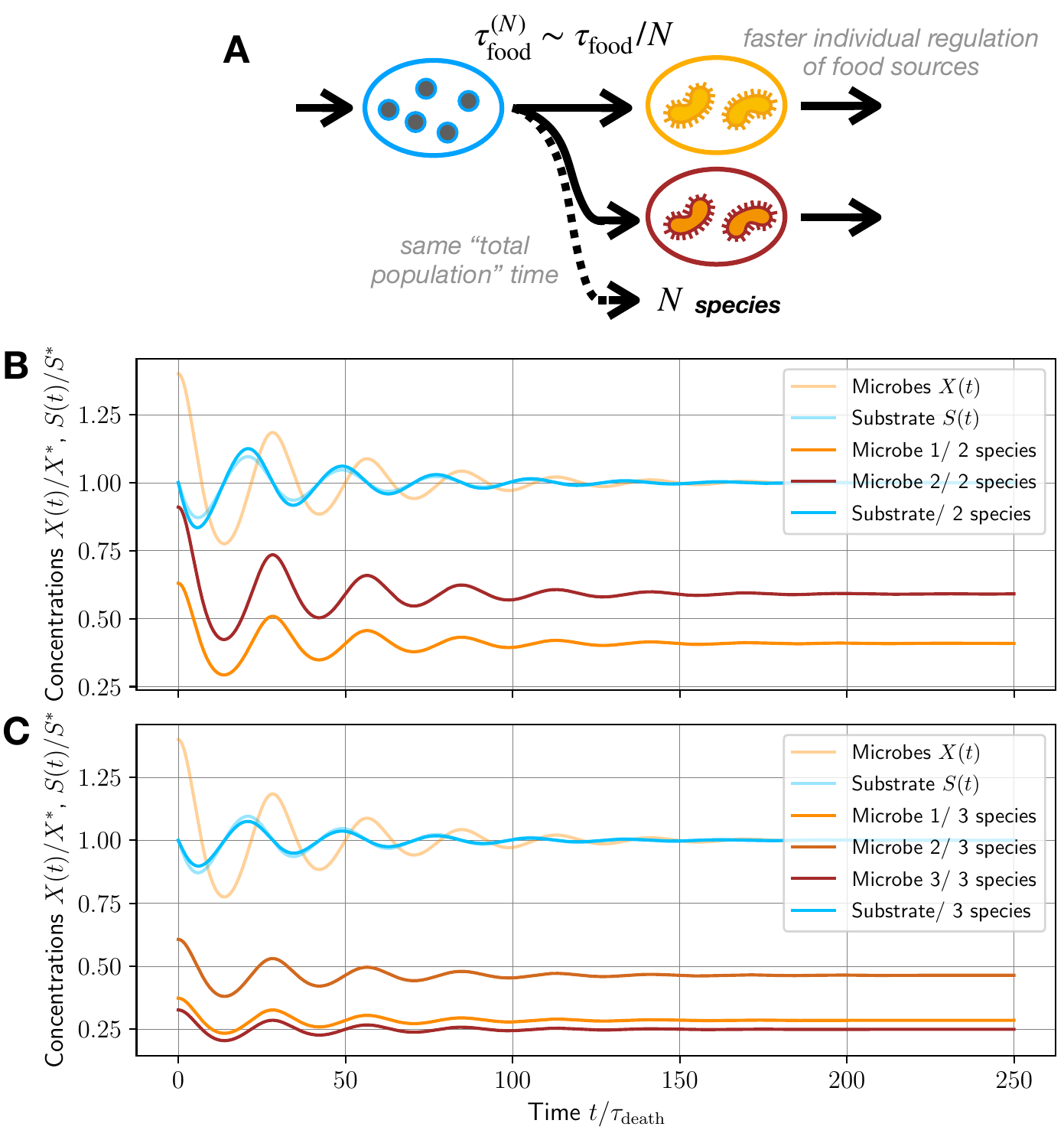}
    \caption{\textbf{Damped oscillatory dynamics around steady state with multiple species.} (A) Schematic of the $N$ species case. (B) and (C) representative solutions around steady state for a system of (B) 2 or (C) 3 species with similar growth and death parameters. In these examples $r = 2 \ /\death$ and $n = 0.1 \ /\death$. }
    \label{fig:figure9}
\end{figure}

In most natural environments, multiple species are found in the same place~\cite{gowda2022genomic}. Supposing they compete for the same food source means we can write the dynamical system of equations as
\begin{equation} \begin{split}
    \dt{\cb_1} &= r \mu_{\cc}(\cc) \cb_1  - \death \, \cb_1  \label{eq:Nspecies} \\
    ... \\
    \dt{\cb_N} &= r \mu_{\cc}(\cc) \cb_N  - \death \, \cb_N \\  
    \dt{\cc} &= \influx \cco -  \gamma  r \mu_\cc(\cc) (\cb_1 + ... + \cb_N)  \\
\end{split}\end{equation} 
where for simplicity we assumed all species have similar growth and death parameters (see also sketch in Fig.~\ref{fig:figure9}-A). We notice that by simply taking $\cb_{\rm tot} = \cb_1 + ... + \cb_N$ we recover the same equations as in the reference case but now for the system of variables $(\cb_{\rm tot},S)$. There is thus no difference in the range of parameters where oscillations arise with multiple species. More precisely, one can see that 
by definition in Eq.~\eqref{eq:tauc}, for species 1, the timescale of adaptation to changes in food abundance is
\begin{equation}
    \tauc^{(N)} = \left( \frac{\partial \frac{d\cb_1}{dt}}{\partial \cc} \bigg|_{*}\right)^{-1} = 1/ r \mu'(\ccn^*) \cbn_1^*.
\end{equation}
and $\taud = 1/\death$. If all species are in similar abundance initially then they are still in similar abundance at steady state and $\cbn_1^* \simeq \cbn^*/N$. Then we find $\tauc^{(N)} \simeq \tauc/N$. So for each individual, the timescale for adaptation to changes in food abundance is faster; however, for the population as a whole the effective timescale of adaptation to changes in food abundance is $N \tauc^{(N)} \simeq \tauc$ and hence there is no change in the phase diagram. This phenomemology is apparent in Fig.~\ref{fig:figure9}-B and C, where we plot typical evolutions around the steady state for 2 or 3 species. In these figures, the shape of the damped oscillations is closely related to the case with only 1 species, showing no strong difference. At steady state, we see that the relative abundance of each of the species reflects the relative abundance at the initial state.

This phenomenology would change if some species had a competitive advantage over others, say some for which the growth rate $r$ is larger than others. In that simple model though, then the only fixed points in the system that are nontrivial correspond to 1 species surviving and all the others extinct. We expect the species with the largest growth rate would in most cases overcome the others, and then the phenomenology around steady state is quite similar to that of the reference case with only 1 species. We leave the investigation of multiple species depending on multiple food sources for further work. 

\subsection{Substrate removal}
\label{sec:substrateremoval}

Our starting equations Eq.~\eqref{eq:model1substrate} are very close to that in Ref.~\cite{khatri2012oscillating} but for a substrate removal term which would modify only the equation for the substrate dynamics as  
\begin{equation} \begin{split}
\label{eq:substrate}
    \dt{\cc} = \influx \cco -  \gamma  r \mu(\cc)\cb  - \outflux \cc
\end{split}\end{equation} 
where $\outflux$ is a rate of removal, which could correspond to removal by external flows in the system, as for instance in a chemostat. The additional $\outflux$ term is not straightforward to include in the general framework, so the stability analysis has to be done from scratch -- details in Appendix~\ref{app:substrate}. Compared to our reference case, the steady state biomass concentration is changed and decreases slightly to 
\begin{equation}
\label{eq:fixedSubstrate}
   (\cbn^*_{\rm out},\ccn^*_{\rm out})  = \left( \cbn^* \left[ 1 - \frac{\outflux}{\influx} \ccn^* \right], \ccn^*\right).
\end{equation}
In practice, the amount of available resources for growth is constantly diminished by the substrate outflow, and thus the carrying capacity is lowered. Doing a naive investigation of the system one finds that the timescales become
\begin{equation}
\begin{split}
        \tauc^{(\rm out)} &= \tauc \frac{1}{1 - \frac{\outflux}{\influx} \ccn^* } \\
        \taud^{(\rm out)} &= \taud
        \label{eq:tauout}
\end{split}
\end{equation}
such that the timescale of adaptation to changes in food abundance increases. This is somewhat intuitive since the removal rate increases food ``scarcity'' and thus tends to slow down adaptation to food dynamics. This would mean, according to our general rule Eq.~\eqref{eq:condition}, that the range of parameter space where oscillations occur is then larger. However, the stability analysis shows that the effect is actually opposite. 

\begin{figure}
    \centering
    \includegraphics[width=0.7\linewidth]{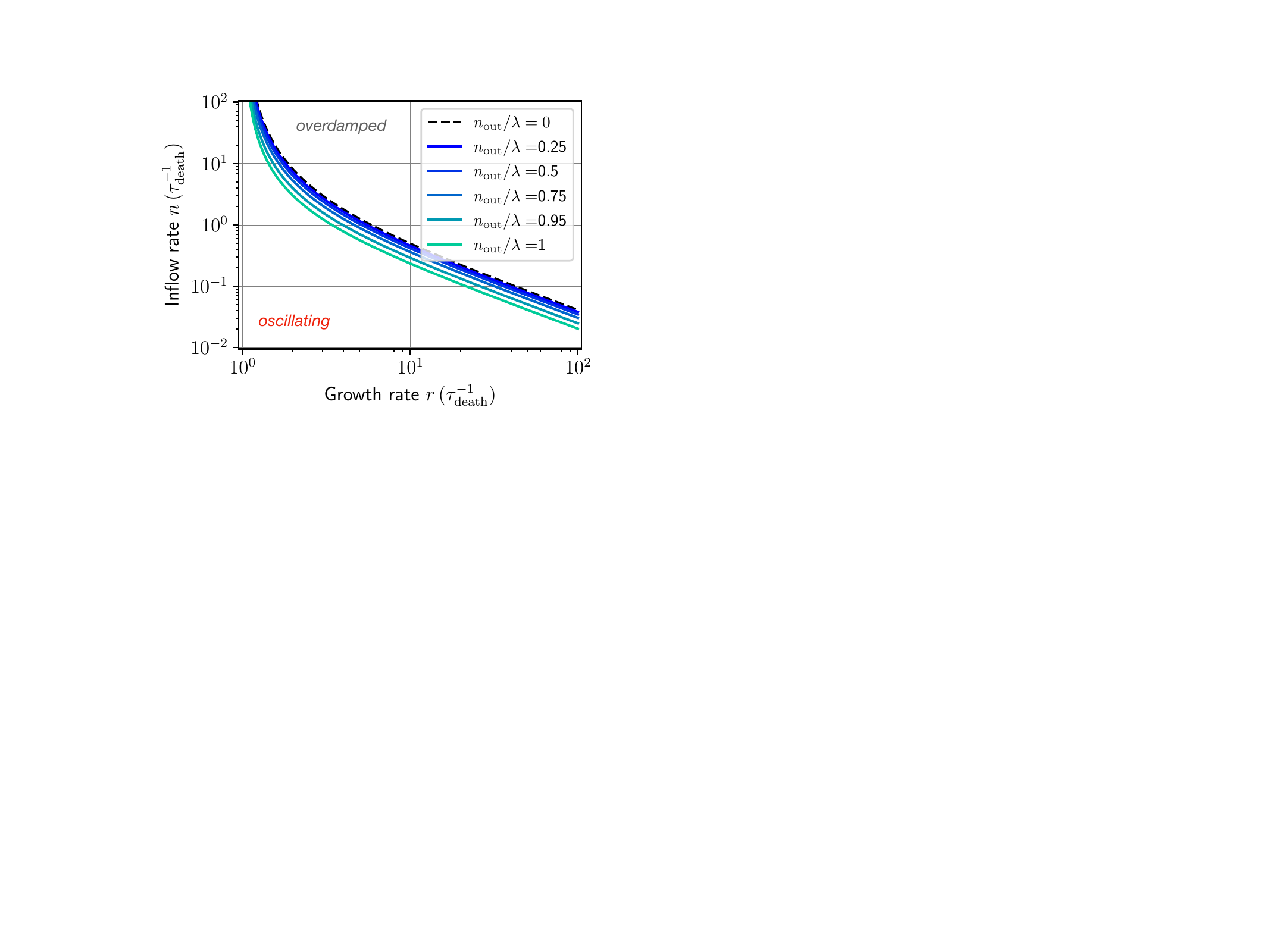}
    \caption{\textbf{Phase diagram in the case of substrate removal.}}
    \label{fig:figure8}
\end{figure}

The stability analysis around the fixed point can be performed and yields slightly more complex results. The eigenvalues around the fixed point are given by 
\begin{equation}
    \frac{\lambda_{\pm}}{\death} = \frac{-\beta_{\rm out} - \frac{\outflux}{\death} \pm \sqrt{\left(\beta_{\rm out} + \frac{\outflux}{\death}\right)^2  - 4 \beta_{\rm out}}}{2}.
\end{equation}
where $\beta_{\rm out} = \taud^{(\rm out)}/\tauc^{(\rm out)}$.
We thus notice that if $\outflux \geq \death$, then there are no oscillations and the dynamics are always overdamped around the fixed point. This framework thus introduces a third timescale in the system, $1/\outflux$.  
In the limit of small outflows, $\outflux \ll \death$ then one can show oscillations occur for values of $\beta_{\rm out} \leq 4 - 2 \outflux/\death$, so over a range that is appears smaller than in the absence of substrate removal. Using Eq.~\eqref{eq:tauout}, we find a condition on the parameters $n, r$ which we report in Fig.~\ref{fig:figure8} for various values of $\outflux/\death$. We find that indeed the range of parameters where oscillations occur is reduced. This analysis shows the limit of our approach based on timescales in the case where external environmental parameters affect the dynamics. To reconcile our approach with these results one should update the definition of the timescale of regulation of food resources as $\taud \simeq (1 + \outflux/2\death)/\death$, which is increased due to removal. 

Note that when $\outflux = \death$ indeed there are no oscillations, this is the ``chemostat'' model~\cite{smith1995theory}. The chemostat model though, prevents to account for special -- yet probably more realistic -- cases where the outflow of different substrates may be different and neglects death or other phenomena that might reduce microbial population.

\subsection{Waste production is not affected by oscillations}



Finally, we investigate the impact of damped oscillations on the production of waste. Waste production, in the form say of carbon dioxide, can be modeled, within our general framework, as
\begin{equation} \begin{split}
    \dt{\cw} &= \zeta \gamma r \mu(\cc)\cb 
    \label{eq:carbonproduction}
\end{split}\end{equation} 
where $\cw$ is the concentration of waste (or carbon dioxide) and $\zeta \leq 1$ is a conversion factor. 
At steady state, $ \dt{\cw} = \zeta \gamma r \mu(\cc^*)\cb^*  = \zeta \influx \cco \equiv r_{\cw}^* \cco$. The waste production rate $ r_{\cw}^* = \zeta \influx $ is directly proportional to the inflow of carbon, as one might expect.

More interestingly, we can study how a solution evolves around a fixed point to quantify the total amount of waste produced before reaching steady state. One can notice from Eq.~\eqref{eq:model1substrate} that
\begin{equation}
    \dt{\cw} = \zeta \left( \influx \cco - \dt{\cc}\right)
\end{equation}
such that the total waste production at long enough times is given by
\begin{equation}
    \cw(t) \simeq \zeta (\cc(0) - \cc^*) +  r_{\cw}^* K_S t.
\end{equation}
The initial substrate concentration thus sets the total waste production. If the initial substrate concentration is larger than the steady state value, substrate in excess will be wasted. Reciprocally, if the initial substrate concentration is lower than the steady state value, inflowing substrate will serve to reach the steady state value and avoid waste. All in all, this yields natural results where if food sources are in excess, waste is generated. This is true regardless of the way the steady state solution is reached, whether in an overdamped or oscillatory way. The presence of oscillations thus does not contribute to increase or decrease waste production. 

\begin{table*}[t]
    \centering
    \begin{tabular}{p{0.15\linewidth}|p{0.25\linewidth}|p{0.25\linewidth}|p{0.2\linewidth}}
        \textbf{Case} & \textbf{Adaptation time to changes in food abundance} & \textbf{Regulation time of food resources} & \textbf{Oscillatory parameter range} \\
        \hline
        Minimal model & $ \tauc = \frac{\death}{r\influx} \left( 1 - \frac{\death}{r} \right)^{-2} $ &$\taud = 1/\death$ &  reference case \\
        Recycling with coefficient $\alpha$ & $\tauc^{(\alpha)} = \tauc(1-\alpha)$ & $\taud^{(\alpha)} = \taud/(1-\alpha)$ & \textit{reduced} \\
        Complementary food sources & $\tauc^{(\rm comp)} = \tauc \frac{2 \left(r/\death -1 \right)^2}{\left(2r/\death -1 \right)^2}$& $\taud^{(\rm comp)} = \taud$ & \textit{reduced} \\
        Necessary food sources & $\tauc^{(\rm nec)} = \tauc  \frac{\left( r/\lambda -1 \right)^2}{2\sqrt{r/\lambda} \left(\sqrt{r/\lambda} -1 \right)^2}$ & $\taud^{(\rm comp)} = \taud$ & \textit{increased}  \\
        Multiple species & $N \tauc^{(\rm N)} = \tauc $ & $\taud^{(N)} \simeq \taud$ & \textit{same} \\
        Substrate removal & $\tauc^{(\rm out)} = \tauc/  \left( 1 -  \outflux s^*/n \right)$ & $\taud^{(\rm out)} \simeq \taud (1 + \outflux/2\death)$ & \textit{reduced} \\
    \end{tabular}
    \caption{\textbf{Summary of the results of the different models investigated in this paper}. }
    \label{tab:my_label}
\end{table*}

Waste production is thus dominated by steady state parameters. 
In cases where the dynamics are more complex, waste production will be modified accordingly. For instance, in the case with recycling, waste production is still given by Eq.~\eqref{eq:carbonproduction}. At steady state though
\begin{equation} \begin{split}
    \dt{\cw} = \zeta \gamma r \mu(\cc_{\alpha}^*)\cb_{\alpha}^*  = \zeta \frac{\influx}{1- \alpha} \cco \equiv \frac{r_{\cw}^*}{1-\alpha} \cco.
\end{split}\end{equation} 
Compared to the case without recycling, carbon production is more important at steady state, naturally because the population sustained at steady state is larger and, thus, produces more waste. Similarly, in the case of complementary (respectively necessary) food sources, we would expect the waste production rate to be larger (respectively smaller). 


\section{Conclusion}


Here, we have reported the conditions under which decaying oscillatory dynamics of a single microbial species and of its food source might arise. Our general framework highlights that only light conditions are required to observe oscillatory dynamics. The key point is to have 4 balancing mechanisms to get an imbalance of timescales: microbial growth, microbial death, food consumption, and food inflow. These mechanisms are so common for microbial systems that the presence of oscillatory dynamics is likely even in more complex systems -- provided the influx of food resources is small enough. There are also no strong conditions on the efficiency of reproduction depending on the amount of available food sources $S$ -- \textit{i.e.} the shape of $\mu(S)$ -- to observe oscillations: it should just be a continuous, increasing function of $S$, which is again something expected and quite common. While we have mostly focused here on damped oscillatory behavior as population and food source relax towards the steady state, we have shown this damped oscillatory behavior is correlated with persistent oscillations in a noisy environment. 

We have rationalized that the oscillations occur if and only if there is an imbalance between the time required for the microbial population to adjust to changes in food abundance $\tauc$ and the time to regulate food supplies $\taud$, equal in our case to the time to regulate excess microbial population numbers. Since microbes rely on food sources to grow, microbes represent the ``downstream'' phenomenon. If microbial adaptation is the faster of the two processes, $\tauc \gtrsim \taud$, it will respond too quickly for food sources to follow and oscillations will occur. We found this principle of competing timescales allowed us to rationalize behavior in different models. The range of parameter space where oscillations would occur was reduced when adding many biophysical components to the model: recycling of microbial necromass; complementary or substitutable food sources or continuously removing food sources. In many of these cases, the phenomenon speeds up adaptation to changes in food abundance, and $\tauc$ decreases. In contrast, relying on multiple necessary or essential food sources has the opposite effect, $\tauc$ increases. Multiple species systems do not modify these dynamics. We provide a summary of these behaviors in Table~\ref{tab:my_label}. Regardless of the biophysical component added, damped oscillatory dynamics around the steady state are always observe, and thus such behavior should be quite generic. However, many environmental factors, especially those that tend to facilitate growth, can reduce the range where this behavior occurs.

We recall, that unlike Lotka-Volterra dynamics -- often relying on logistic growth equations -- which investigate only biotic, \textit{i.e.} reproducing species, here, in the consumer-resource (CR) model we investigate both biotic and abiotic, \textit{i.e.} non-reproducing, compounds. CR models can converge to a logistic-growth-like model, assuming a sharp enough separation of timescales between biotic growth and abiotic depletion rate~\cite{reynolds2013can}. This allows us to draw further understanding on the Lotka-Volterra models. In particular, the steady state microbial population of the CR model corresponds to the carrying capacity of a logistic growth model, and the decay rate can be viewed as the logistic growth rate. However, a single microbial species described by a logistic growth equation will not undergo oscillations, and so the oscillatory phenomena described here can only be accounted for with CR models. 



Beyond the homogeneous, simple situation explored here, many extensions of the model are possible, especially in space and time. First, it is expected that oscillations originate at different scales, especially at smaller scales. Even at the level of a single cell, oscillations in the abundance of a given substrate can undergo large fluctuations and oscillations~\cite{bi2023dynamic,schaefer1999automated}. This suggests to make models with even more layers of scales, and rigorously coarse-grain models to obtain effective behavior at larger scales~\cite{reynolds2013can,skichko2006mathematical}, beyond standard Lotka-Volterra or logistic growth models. Second, naturally at these smaller scales fluctuations occur because of the small amount of species involved, which is related to the possibility of extinction.  
Finally, most ecosystems reside in highly heterogeneous environments. For instance, the distribution of nutrients in soils is highly heterogeneous starting with oxygen content~\cite{angle2017methanogenesis,fenchel2008oxygen} necessarily shaping microbial communities. There is thus a vast interplay of dynamics to be understood at the nutrient-microbiome level in more realistic systems.







\section*{Data availability}

All data needed to evaluate the conclusions in the paper are present in the paper. 

\section*{Author Declarations}

\subsection*{Conflict of Interest}

The authors have no conflicts to disclose.

\subsection*{Author Contributions}

\textbf{Benedetta Ciarmoli}: Investigation (equal); Software (equal); Visualization (equal); Writing - original draft (equal). \textbf{Sophie Marbach}: Conceptualization (lead); Investigation (equal); Methodology (lead); Supervision (lead); Visualization (equal); Writing - original draft (equal); Writing - review \& editing (lead).

\section*{Acknowledgements}

We wish to acknowledge fruitful discussions with Seppe Kuehn and Nelly Henry. This research project was supported by the Institut de la Transition Environnementale of the Alliance Sorbonne Universit\'e.

\appendix


\section{Calculation details for the case of complementary food sources}
\label{app:complementary}

Starting from Eq.~\eqref{eq:complementary}, the Jacobi matrix around the nontrivial fixed point is 
\begin{equation}
    J = \begin{pmatrix}
        0 & \beta_{\rm comp} & \beta_{\rm comp} \\ 
        -1/2 &- \beta_{\rm comp} & 0 \\
        -1/2 & 0 & -\beta_{\rm comp}
    \end{pmatrix}.
\end{equation}
One can show the eigenvalues of this matrix are given by $\tilde{\lambda}_0 = -\beta_{\rm comp}$ and 
\begin{equation}
    \tilde{\lambda}_{\pm} = \frac{-\beta_{\rm comp} \pm \sqrt{\beta_{\rm comp}^2  - 4 \beta_{\rm comp}}}{2}.
\end{equation}
The solutions of the system around the fixed point will exhibit a damped oscillatory behavior if and only if $\beta_{\rm comp} \leq 4$. 

\section{Calculation details for the case of necessary food sources}
\label{app:nec}

Starting from Eq.~\eqref{eq:modelcomplementarysimplified}, the Jacobi matrix around the nontrivial fixed point is 
\begin{equation}
    J = \begin{pmatrix}
        0 & \beta_{\rm nec} & \beta_{\rm nec} \\ 
        -1 &- \beta_{\rm nec} & - \beta_{\rm nec} \\
        -1 & - \beta_{\rm nec} & -\beta_{\rm nec}
    \end{pmatrix}.
\end{equation}
One can show the eigenvalues of this matrix are given by $\tilde{\lambda}_0 = 0$ and 
\begin{equation}
    \tilde{\lambda}_{\pm} =-\beta_{\rm nec} \pm \sqrt{\beta_{\rm nec}^2  - 2 \beta_{\rm nec}}.
\end{equation}
The solutions of the system around the fixed point will exhibit a damped oscillatory behavior if and only if $\beta_{\rm nec} \leq 2$. 

\section{Calculation details for the case of substrate removal}
\label{app:substrate}

Starting with Eqs.~\eqref{eq:model1substrate} and \eqref{eq:substrate}, the jacobian matrix around the non-trivial fixed point Eq.~\eqref{eq:fixedSubstrate} is 
\begin{equation}
    J = \begin{pmatrix}
        0 &  \frac{r}{\death} \mu'(\ccn^*_{\rm out})\cbn_{\rm out}^* \\
        -\frac{r}{\death} \mu(\ccn_{\rm out}^*) & - \frac{r}{\death} \mu'(\ccn_{\rm out}^*)\cbn_{\rm out}^* -  \frac{\outflux}{\death}
    \end{pmatrix} 
\end{equation}
and simplifies to
 \begin{equation}
    J = \begin{pmatrix}
        0 & \beta_{\rm out} \\
         - 1 & - \beta_{\rm out} -  \frac{\outflux}{\death}
    \end{pmatrix}
\end{equation}
where the definition of the non-dimensional parameter $\beta_{\rm out} =\frac{r}{\death} \mu'(\ccn^*_{\rm out})\cbn_{\rm out}^*$. Let $\tilde{\lambda}_\pm = \lambda_{\pm}/\lambda$ be the nondimensional eigenvalues of $J$, they satisfy
\begin{equation}
    \frac{\lambda_{\pm}}{\death} = \frac{-\beta_{\rm out} - \frac{\outflux}{\death} \pm \sqrt{\left(\beta_{\rm out} + \frac{\outflux}{\death}\right)^2  - 4 \beta_{\rm out}}}{2}.
\end{equation}
These eigenvalues lead to damped oscillatory behavior if and only if $\left(\beta_{\rm out} + \frac{\outflux}{\death}\right)^2 - 4 \beta_{\rm out} \leq 0$ which defines new boundaries for the oscillatory behavior in phase space. This corresponds to a limiting equation for $\beta_{\rm out}$ as 
\begin{equation}
    \beta_{\rm out} \leq 2 + 2 \sqrt{1 - \frac{\outflux}{\death}} - \frac{\outflux}{\death} \simeq 4 - 2 \frac{\outflux}{\death}
\end{equation}
where the last equality is obtained in the limit when $\outflux \ll \death$. Keeping the ratio between $\outflux$ and $\death$ arbitrary, this corresponds to a boundary equation for the influx rate $\influx$ as 
\begin{equation}
    \frac{\influx}{\death} \leq \frac{\death}{r \mu'(\ccn^*)}\left(2 \sqrt{1 - \frac{\outflux}{\death}} + 2 + \frac{\outflux}{\death} \left( \frac{r}{\lambda}\mu'(\ccn^*) \ccn^* - 1 \right) \right)
\end{equation}
which was used to plot the phase diagrams in Fig.~\ref{fig:figure8}.
When $\outflux \ll \death$ this latter equation simplifies to 
\begin{equation}
    \frac{\influx}{\death} \lesssim \frac{\death}{r \mu'(\ccn^*)}\left(4   - \frac{\outflux}{\death} (2 - \frac{r}{\lambda}\mu'(\ccn^*)) \ccn^*  \right).
\end{equation}


%



\end{document}